\documentclass[prd,showpacs,nofootinbib,preprint, 10 pt]{revtex4}
\usepackage[a4paper, total={7in, 10in}]{geometry}
\usepackage{amsmath,graphicx,xcolor,epsfig,slashed}
\usepackage{appendix}
\usepackage{tabularx}
\usepackage{float}
\usepackage{pdfpages}
\usepackage{epstopdf}
\usepackage{subfigure}
\usepackage{caption}
\usepackage{multirow}
\usepackage{booktabs,siunitx}
\usepackage{hyperref}
\usepackage{orcidlink}

\begin{document}
\hfill DESY 24-061 \\[5mm]
\begin{center}
{\bf\Large\boldmath
Signature decay modes of the compact doubly-heavy tetraquarks
$T_{bb\bar{u}\bar{d}}$ and $T_{bc\bar{u}\bar{d}}$}\\[5mm]
\par\end{center}

\begin{center}
\setlength{\baselineskip}{0.2in}
Ahmed Ali\,\orcidlink{0000-0002-1939-1545}, \\
Deutsches Elektronen-Synchrotron DESY, Notkestr. 85, 22607 Hamburg,Germany\\
Ishtiaq Ahmed \,\orcidlink{0000-0001-7583-5369}, \\
National Center for Physics, Quaid-i-Azam University Campus, Islamabad, 44000, Pakistan\\
Muhammad Jamil Aslam \,\orcidlink{0000-0001-9192-066X}, \\
Department of Physics, Quaid-i-Azam University, Islamabad, 45320, Pakistan
\end{center}
\textbf{Abstract}\\[5mm] 
Based on the expectations that the lowest-lying doubly-bottom tetraquark
$T_{bb\bar u \bar d}$ ($J^P = 1^+$) and the bottom-charm tetraquark 
$T_{bc\bar u \bar d}$ ($J^P = 0^+$) are stable against strong and electromagnetic decays, 
we work out a number of semileptonic and non-leptonic weak decays of these hadrons, 
making use of the heavy quark symmetry. In doing this, we concentrate on the exclusive decays 
involving also tetraquarks in the final states, i.e., transitions such as 
$T_{bb\bar u \bar d} \to T_{bc\bar u \bar d}\, (\ell^- \nu_\ell,\, h^-)$ and 
$T_{bc\bar u \bar d} \to T_{cc\bar u \bar d}\, (\ell^- \nu_\ell,\, h^-)$, where 
$h^- = \pi^-, \rho^-, a_1^-, D^-_s, D^{*-}_s$. So far, only the $J^P = 1^+$ tetraquark 
$T_{cc\bar u \bar d}$ has been discovered, which we identify with the $I = 0$ $T_{cc}^+$ object, 
compatible with $J^P = 1^+$ and having the 
pole mass relative to the $D^{*+} D^0$ mass threshold and decay widths 
$\delta m  = M (T_{cc}^+) - ( M (D^{*+}) + M (D^0) ) = - 360 \pm 40^{+4}_{-0}$ keV
and $\Gamma (T_{cc}^+) = 48^{+2}_{-14}$ keV.  
Experimental discoveries of the transitions worked out here, and related ones involving 
doubly-heavy baryons, will quantify the diquark-antidiquark component of these tetraquarks.


\section{Introduction}\label{sec:1}
The physics of the multiquark hadrons has emerged as a recurrent theme
in high-energy physics, in particular, at the LHC.
Prominent among these hadrons is the doubly-charm tetraquark $T^+_{cc\bar u \bar d}$, 
called $T_{cc} (3875)^+$, discovered by LHCb~\cite{LHCb:2021vvq} in prompt proton-proton collisions 
as a very narrow state (${\rm FWHM} = 48$ keV) in the final state $D^0 D^0 \pi^+$. 
It is found just below the $D^{*+} D^0$ threshold, has a mass 
$m (T_{cc \bar u \bar d}^+) = 3874.83 \pm 0.11$ MeV~\cite{ParticleDataGroup:2022pth}, 
and is compatible with being an isoscalar ($I = 0$) with the spin-parity $J^P = 1^+$. 
Its characteristic size, calculated from the binding energy $\Delta E = -360 \pm 40 ^{+4}_{-0}$ keV, 
yields $R_{\Delta E} = 7.49\pm 0.42$ fm, too large for a compact hadron,
nevertheless having a sizeable cross-section, estimated as
$\sigma (pp \to T_{cc\bar u\bar d} + X) = (45 \pm 20)$ nb at $\sqrt s = 13$ TeV for the typical
LHCb acceptance ($2 < p_T < 20~{\rm GeV};\, 2 < y < 4.5)$~\cite{LHCb:2021auc}. 
Its nature (a hadron molecule or compact tetraquark) is still being debated.

\vspace*{3mm}
With this discovery, as well as that of the doubly-charm ($C = 2$) baryon $\Xi_{cc}^{++}$, 
having the quark content $(ucc)$~\cite{LHCb:2017iph}, the focus of the experimental and
theoretical research is now on their heavier counterparts, the bottom-charm hadrons 
($C = 1,\,  B = - 1$), $\Xi_{bcq}$ and $T_{bc\bar u \bar d}$, and eventually on the doubly-bottom 
( $B = -2$) baryons $\Xi_{bbq}$, $(q = u,\, d,\, s)$, and the doubly-bottom tetraquarks 
$T_{bb\bar u \bar d}$, $T_{bb\bar u \bar s}$, and $T_{bb\bar d \bar s}$. 
Theoretically, stable (w.r.t. strong decays) heavy multiquark states were predicted a long time 
ago~\cite{Ader:1981db,Manohar:1992nd,Silvestre-Brac:1993zem,Brink:1998as,Ebert:2007rn}.
Their phenomenology has been studied in several competing theoretical approaches. 
Among other frameworks, the ones based on diquarks~\cite{Jaffe:1976ig,Jaffe:2003sg} 
have been extensively used to model compact multiquark states~\cite{Ali:2019roi}. 
More recently, heavy quark~--- heavy diquark symmetry has been invoked to relate singly heavy mesons, 
anti-baryons, and doubly-heavy baryons and tetraquarks~\cite{Mehen:2017nrh,Eichten:2017ffp}. 
Lately, the role of the local diquark-antidiquark operators in the spectroscopy of tetraquark 
hadrons has also been investigated using the Lattice-QCD framework~\cite{%
Bicudo:2021qxj,Alexandrou:2023cqg,Radhakrishnan:2024ihu,Alexandrou:2024iwi}. 
In particular, the masses of the ground-state doubly-heavy tetraquarks have been calculated 
using both the meson-meson and the diquark-antidiquark operators. In one such study, carried out
for the $T_{bb\bar u \bar d}$ and $T_{bb\bar u \bar s}$ tetraquarks, an approximately even mix 
of meson-meson and diquark-antidiquark component are found~\cite{Bicudo:2021qxj}. Further studies 
are needed to quantify this prediction, and experimental proofs of the compact nature of the multiquark 
states unambiguously are required. We argue here that the weak transitions of doubly-heavy tetraquarks 
may also reveal the existence of heavy diquarks (and antidiquarks), and we work out some characteristic 
decays reflecting the underlying compact structures.

\vspace*{3mm}
We note that the lowest-mass doubly-bottom tetraquarks $T_{bb\bar u \bar d}$ and $T_{bb\bar u \bar s}$ 
are estimated to lie below their respective strong-decay thresholds. A recent Lattice-QCD 
simulation~\cite{Alexandrou:2024iwi} puts their masses (measured w.r.t. their respective 
thresholds) as $\delta m = -100 \pm 10^{+36}_{-43}$ MeV for the $T_{bb\bar u \bar d}$ and 
$\delta m = -30 \pm 3^{+11}_{-31}$ MeV for the $T_{bb\bar u \bar s}$, similar to earlier 
estimates~\cite{Alexandrou:2024iwi-Fig-13}. Some weak decays of these tetraquarks have been 
worked out in a number of papers~\cite{Ali:2018ifm,Agaev:2020dba,Agaev:2020mqq,Hernandez:2019eox}, 
providing estimates for the lifetimes and branching ratios. Of particular interest are the
inclusive decays of doubly-bottom tetraquarks, $T_{bb \bar u \bar d} \to B_c^{(*)} + X$, 
as well as of the corresponding baryons~\cite{Kiselev:2001fw,Ridgway:2019zks}, $\Xi_{bbq} \to B_c^{(*)} + X$,
with detached $B_c$-vertex~\cite{Gershon:2018gda} due to the predicted long lifetimes of these hadrons, 
estimated to lie in the range $0.4 - 0.8$ ps~\cite{Ali:2018ifm,Kiselev:2001fw,Karliner:2014gca,Berezhnoy:2018bde}. 

\vspace*{3mm}
Until recently, there was no consensus on the issue of whether the lowest-mass ($C = 1$, $B = -1$) 
tetraquark state is stable against strong and electromagnetic decays, as well as on the assignment 
of the quantum numbers~\cite{Radhakrishnan:2024ihu-Fig-8}. However, two recent Lattice-based 
estimates~\cite{Radhakrishnan:2024ihu,Alexandrou:2023cqg} have posted the mass of the lowest-lying state, 
with spin-parity $J^P = 0^+$, below the $D B$-threshold. If confirmed by further theoretical developments, 
this would make a strong case for the $T_{bc \bar u \bar d}$ tetraquark as the first multiquark 
state to decay weakly. The corresponding $J^P = 1^+$ tetraquark is also found in these 
studies~\cite{Radhakrishnan:2024ihu,Alexandrou:2023cqg} to be below the $D B^*$-threshold, though it would 
decay via electromagnetic transition to $D B \gamma$ or radiatively to the lower-mass $J^P = 0^+$ state. 
This information is helpful for the experimental searches of $T_{bc \bar u \bar d}$ tetraquark 
in focusing on the final states, which can only be reached by charged current weak interactions. 

\vspace*{3mm}
With this hindsight, we shall concentrate here on the weak decays of the doubly-bottom $T_{bb \bar u \bar d}$ 
and the bottom-charm $T_{bc \bar u \bar d}$ tetraquarks in which the meson-meson and the diquark-antidiquark 
components are established on the Lattice~\cite{Bicudo:2021qxj}, and also in the Born-Oppemheimer 
approximation in QCD with heavy quarks $(Q Q)$ and light antiquarks $(\bar u \bar d)$~\cite{Maiani:2022qze}.
Weak decays from the meson-meson components (such as $B B^{(*)}$, $D B^{(*)}$) follow the known patterns, 
well documented in the Particle Data Group tables~\cite{ParticleDataGroup:2022pth} and are included 
in the current experimental search strategies~\cite{Blusk:21,Polyakov:23}. The ones, following from 
the diquark-antidiquark component discussed in this Letter are new, or at least have not been studied 
so far quantitatively. In particular, they lead to weak decays involving tetraquarks both in the initial 
and final states, i.e., they induce transitions, such as $T_{bb \bar u \bar d} \to T_{bc \bar u \bar d} + X$
and $T_{bc \bar u \bar d} \to T_{cc \bar u \bar d} + X$. We work out some exclusive decays, 
$T_{bb \bar u \bar d} \to T_{bc \bar u \bar d}\, (\ell^- \nu_\ell,\, h^-)$ and 
$T_{bc \bar u \bar d} \to T_{cc \bar u \bar d}\, (\ell^- \nu_\ell,\, h^-)$, where 
$\ell = e,\, \mu,\, \tau$ and $h^- = \pi^-, \rho^-, a_1^-, D_s^-, D_s^{*-}$~\cite{Maiani:LHCb2023}. 
Since doubly-heavy tetraquarks and the doubly-heavy baryons are related by heavy diquark symmetry,
we also anticipate decays such as $T_{bb \bar u \bar d} \to \Xi_{bc}^+\, \bar p\; (\ell^- \nu_\ell)$, 
$T_{bb \bar u \bar d} \to \Xi_{bc}^+\, \bar p\; (\pi^-, \rho^-, a_1^-, D_s^-, D_s^{*-})$, 
$T_{bc \bar u \bar d} \to \Xi_{cc}^{++}\, \bar p\; (\ell^- \nu_\ell)$, and
$T_{bc \bar u \bar d} \to \Xi_{cc}^{++}\, \bar p\; (\pi^-, \rho^-, a_1^-, D_s^-, D_s^{*-})$. 
The Feynman diagrams inducing these decays are similar to the ones shown below for the 
$T_{bb \bar u \bar d} \to T_{bc \bar u \bar d} + X$ and $T_{bc \bar u \bar d} \to T_{cc \bar u \bar d} + X$ 
decays, except that they require an additional $q \bar q$-pair production from the vaccum. 
Together they represent signature decay modes of compact tetraquarks, reflecting the heavy diquark 
configurations in their wave-functions, which dominate over the mesonic configurations at short 
inter-heavy-quark distances $r$~\cite{Bicudo:2021qxj,Maiani:2022qze}. For large values of $r$, 
the tetraquark $T_{bb \bar u \bar d}$ wave-function is essentially determined by the asymptotic 
states $BB$ and $BB^*$. Likewise, for the  $T_{bc \bar u \bar d}$, the asymptotic states are 
$DB$ and $DB^*$, which would decay independently, giving rise dominantly to multi-body decays. 
On the other hand, diquark part of the wave-function would lead to two-body non-leptonic decays, 
such as $T_{bb \bar u \bar d} \to T_{bc \bar u \bar d} + (\pi^-, \rho^-, a_1^-, D_s^-, D_s^{*-})$ 
and $T_{bc \bar u \bar d} \to T_{cc\bar u \bar d} + (\pi^-, \rho^-, a_1^-, D_s^-, D_s^{*-})$ 
with measurable branching fractions, not anticipated from the corresponding meson-meson components. 
 
\vspace*{3mm}
Since there is no annihilation or $W$-exchange diagrams allowed in these transitions, these decays 
take place via the so-called color-allowed tree diagrams, shown 
in Figs.~\ref{FeynDiag1} and~\ref{FeynDiag2}. Here, $b$-quark acts as the active (or valence) quark. 
There are two of them in the doubly-bottom tetraquarks $T_{bb \bar u\bar d}$ and one in $T_{bc \bar u \bar d}$. 
Weak interaction induces the (dominant) $b \to c$ transition, but the crucial difference is that 
the $bb$-diquark emerges in the weak decays as an intact $bc$-diquark. Invoking the heavy diquark~--- 
heavy quark symmetry, we relate weak decays of these tetraquarks to the corresponding $B \to (D,\, D^*)$ 
weak decays of the $B$-mesons. The decay rates are worked out in the Heavy Quark Symmetry limit, using 
the Heavy Quark Effective Theory (HQET) framework. The reported branching ratios are encouraging 
to be measured. We note that the High-Luminosity LHC (HL-LHC) at CERN~\cite{Azzi:2019yne,Cerri:2018ypt}, 
and the planned Tera-$Z$ factories~\cite{FCC:2018evy,CEPCStudyGroup:2018ghi} are estimated to have 
large enough data samples~\cite{Ali:2018ifm,Ali:2018xfq} to carry out the required measurements.

\vspace*{3mm}
In Section~\ref{sec:2}, we calculate the semileptonic decays $T_{bb \bar u \bar d} \left ( J^P = 1^+ \right ) 
\to T_{bc \bar u \bar d} \left ( J^P = 0^+ \right ) \ell^-\nu_\ell$. The decay $T_{bb \bar u \bar d} 
\left ( J^P = 1^+ \right ) \to T_{bc \bar u \bar d} \left ( J^P = 1^+ \right ) \ell^- \nu_\ell$ are 
presented in Section~\ref{sec:3}. Non-leptonic decays $T_{bb \bar u \bar d} \to T_{bc \bar u \bar d}\, h^-$ 
($h^- = \pi^-, \rho^-, a_1^-, D^-_s, D^{*-}_s$) are discussed in Section~\ref{sec:4}. The semileptonic 
and non-leptonic decays $T_{bc \bar u \bar d} \left ( J^P = 0^+ \right ) \to T_{cc \bar u \bar d} 
\left ( J^P = 1^+ \right ) (\ell^- \nu_\ell,\, h^-)$ are discussed in Section~\ref{sec:5}. We conclude 
with a summary in Section~\ref{sec:6}. 

\section{Semileptonic decays $T_{bb \bar u \bar d} \left ( J^P = 1^+ \right ) \to 
T_{bc \bar u \bar d} \left ( J^P = 0^+ \right ) \ell^- \nu_\ell$}
\label{sec:2}

\begin{figure}[!htb]
\centering\scalebox{1}{
\begin{tabular}{cc}
\includegraphics[width=2.5in,height=1.8in]{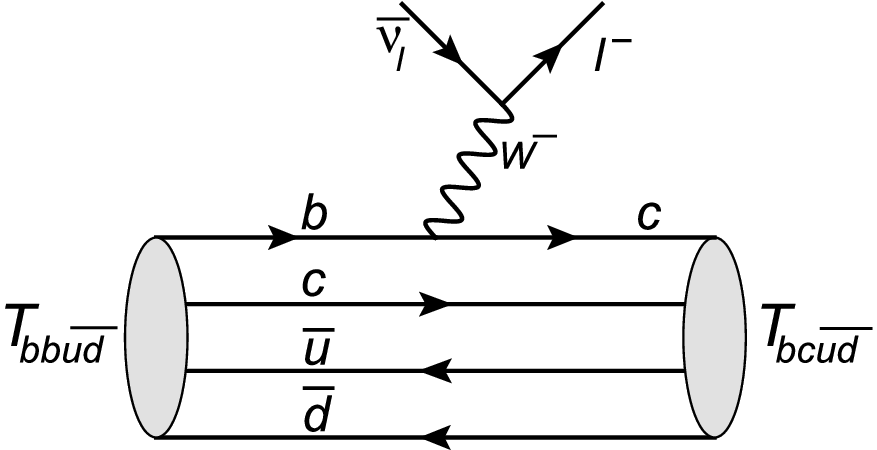}\hspace{0.5cm}\qquad & \qquad
\includegraphics[width=2.5in,height=1.8in]{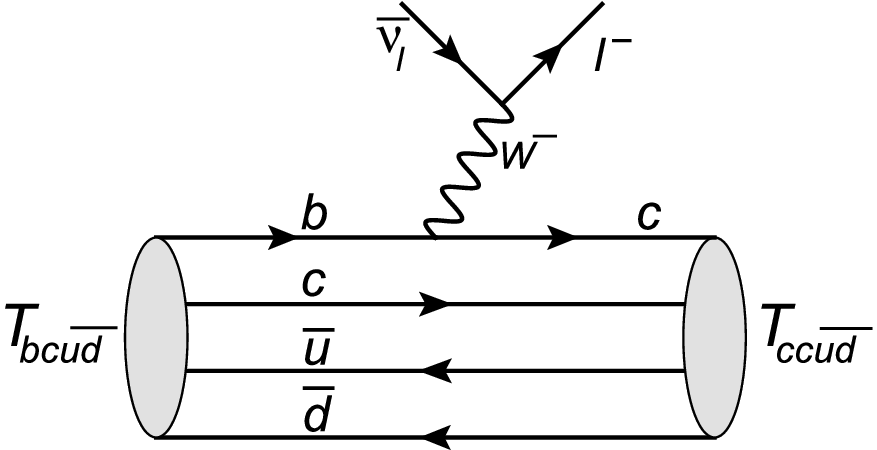} \\
(a)\hspace{0.5cm}\qquad &  \qquad (b)
\end{tabular}}
\caption{
Feynman diagram for (a) $T_{bb\bar u \bar d} \to T_{bc\bar u \bar d}\, \ell^- \nu_\ell$, 
and (b) $T_{bc\bar u \bar d} \to T_{cc\bar u \bar d}\, \ell^- \nu_\ell$, where $\ell = e,\, \mu,\, \tau$.
}
\label{FeynDiag1}
\end{figure}

The decay $T_{bb \bar u \bar d} \to T_{bc \bar u \bar d} \ell^- \nu_\ell$ is governed 
by the following effective Hamiltonian: 
\begin{equation}
\mathcal{H}_{\text{eff}} = 4\, \frac{G_F}{\sqrt 2}\, V_{cb} 
\left ( \bar c \gamma^\mu P_L b \right ) \left ( \bar\ell \gamma_\mu P_L \nu_\ell \right ) , 
\label{eq:06}
\end{equation}
where $P_L = (1 - \gamma_5)/2$, and the corresponding Feynman diagram is shown in Fig.~\ref{FeynDiag1}(a).

\vspace*{3mm}
As discussed in the Introduction, we assume that the initial state tetraquark, $T_{bb \bar u \bar d}$, has 
the spin-parity $J^P =1^+$ (axial-vector) and the final state $T_{bc \bar u \bar d}$ has $J^P = 0^+$ (scalar). 

\vspace*{3mm}
We shall use the Heavy Quark Effective Theory (HQET)~\cite{Manohar:2000dt,Neubert:1993mb}
to calculate the matrix elements $\left \langle T_{bc \bar u \bar d} \left ( p^\prime \right) \left | 
\bar c \gamma^\mu b \right | T_{bb \bar u \bar d} \left (p, \varepsilon \right ) \right\rangle$ and 
$\left\langle T_{bc \bar u \bar d} \left ( p^\prime \right ) \left | \bar c \gamma^\mu \gamma_ 5 b \right | 
T_{bb \bar u \bar d} \left ( p, \varepsilon \right ) \right\rangle$. To that end, we recall that
the superfield combining the pseudo-scalar and vector mesons in HQET can be written as follows: 
\begin{equation}
\mathcal{H}_v^{\left ( Q \right)} = \frac{1 + \slashed{v}}{2} 
\left [ P_v^{* \left ( Q \right ) \mu} \gamma_\mu - P_v^{\left ( Q \right )} \gamma_5 \right ], 
\label{eq:01}
\end{equation}
where $Q = b,\, c$,  $v^\mu$ is the meson 4-velocity ($v^2 = 1$), 
and for the vector field $\left ( v \cdot P_v^{* \left ( Q \right )} \right ) = 0$.  
Similarly, for the axial-vector and the scalar fields, we just have to multiply it with $\gamma_5$ 
from the right, i.e., 
\begin{equation}
\mathcal{H}_v^{\left ( Q \right )} = \frac{1 + \slashed{v}}{2} 
\left [ P_v^{* \left ( Q \right ) \mu} \gamma_\mu \gamma_5 - P_v^{\left ( Q \right )} \right ], 
\label{eq:02}
\end{equation}
where the axial-vector and scalar parts are   
\begin{equation}
\mathcal{H}_{A, v}^{\left ( Q \right )} = \frac{1 + \slashed{v}}{2} 
\left [ P_v^{* \left ( Q \right ) \mu} \gamma_\mu \gamma_5 \right ] , 
\quad 
\mathcal{H}_{S, v}^{\left ( Q \right )} = - \frac{1 + \slashed{v}}{2} \, P_v^{\left ( Q \right )} . 
\label{eq:03}
\end{equation}
The corresponding Dirac-conjugate field can be obtained as
\begin{align}
\bar{\mathcal{H}}_{A, v}^{\left ( Q \right )} & = 
\gamma^0 \mathcal{H}_{A, v}^{\left ( Q \right ) \dagger} \gamma^0 = 
\left [ P_v^{* \left ( Q \right ) \dagger \mu} \gamma^0 \gamma_5 \gamma_\mu^\dagger \right ] 
\frac{1 + \slashed{v}^\dagger}{2} \gamma^0 
\nonumber \\
& = - \left [ P_v^{* \left ( Q \right ) \dagger \mu} \gamma_5 \gamma^0 \gamma_\mu^\dagger \right ] 
\gamma^0 \gamma^0 \frac{1 + \slashed{v}^\dagger}{2} \gamma^0  
= \left [ P_v^{* \left ( Q \right ) \dagger \mu} \gamma_\mu \gamma_5 \right ] \frac{1 + \slashed{v}}{2}, 
\label{eq:04}
\end{align}
since $\gamma^0 \gamma_\mu^\dagger \gamma^0 = \gamma_\mu$ and $\left \{ \gamma_\mu, \gamma_5 \right \} = 0$. 
The important properties are 
\begin{equation}
\slashed{v} \mathcal{H}_{A, v}^{\left ( Q \right )} = \mathcal{H}_{A, v}^{\left ( Q \right )}, 
\quad 
\mathcal{H}_{A, v}^{\left ( Q \right )} \slashed{v} = \mathcal{H}_{A, v}^{\left ( Q \right )}. 
\label{eq:05}
\end{equation}
The Dirac-conjugate of the scalar part can be written as
\begin{equation}
\bar{\mathcal{H}}_{S, v}^{\left ( Q \right )} = 
\gamma^0 \frac{1 + \slashed{v}^\dagger}{2} P_v^{\left ( Q \right ) \dagger} \gamma^0 = 
\frac{1 + \slashed{v}}{2}\, P_v^{\left ( Q \right ) \dagger}. 
\label{eq:04a}
\end{equation}

In the HQET, the matrix elements can be parameterized as
\begin{align}
\frac{\left\langle T_{bc\bar{u}\bar{d}}\left(v^{\prime}\right)\left|\bar{c}\gamma^{\mu}b\right|T_{bb\bar{u}\bar{d}}\left(v,\varepsilon\right)\right\rangle }{\sqrt{m_{T}m_{T^{\prime}}}} & = h_1 \left(w\right)i\varepsilon^{\mu\nu\alpha\beta}\varepsilon_{\nu}v_{\alpha}v_{\beta}^{\prime},\nonumber \\
\frac{\left\langle T_{bc\bar{u}\bar{d}}\left(v^{\prime}\right)\left|\bar{c}\gamma^{\mu}\gamma_{5}b\right|T_{bb\bar{u}\bar{d}}\left(v,\varepsilon\right)\right\rangle }{\sqrt{m_{T}m_{T^{\prime}}}} & = h_2 \left(w\right)\left(w+1\right)\varepsilon^{\mu}-h_{3}\left(w\right)\left(\varepsilon\cdot v^{\prime}\right)v^{\mu}-h_{4}\left(w\right)\left(\varepsilon\cdot v^\prime\right) v^{\prime\mu}.
\label{eq:07}
\end{align}
where $m_T$ and $m_{T^\prime}$ denote the masses of the initial $\left ( T_{bb \bar u \bar d} \right )$
and final $\left ( T_{bc \bar u \bar d} \right )$ tetraquark states, respectively. The corresponding 
form factors for the transitions $B \to (D,\, D^*) \ell^- \nu_\ell$ have been investigated at length,
but the ones for the tetraquarks are yet to be studied.

\vspace*{3mm}
Using the heavy-quark spin symmetry to relate the various form factors, 
at the leading order (LO) in the heavy quark mass, we can write
\begin{equation}
\left\langle \mathcal{H}_{S, v^\prime}^{\left ( Q^\prime \right )} \left | 
\bar Q^\prime \Gamma Q \right | \mathcal{H}_{A, v}^{\left ( Q \right )} \right\rangle 
= - \xi \left ( w \right ) \text{Tr} \left\{ 
\bar{\mathcal{H}}_{S, v}^{\left ( Q^\prime \right )} \Gamma\mathcal{H}_{A, v}^{\left ( Q \right )} \right\}, 
\label{eq:08}
\end{equation}
where $\Gamma$ is a particular combination of $\gamma$-matrices and $Q^\prime$ and $Q$ 
are the $c$- and $b$-quarks with velocities $v^\prime$ and $v$, respectively. It involves 
a single non-perturbative function $\xi \left ( w \equiv v \cdot v^\prime \right )$, i.e., 
the Isgur-Wise function. We note that there are two $b$-quarks present in $T_{bb \bar u \bar d}$, 
and in the weak transition only one active $b$-quark is involved at a time, with the other 
treated as a spectator. There are no annihilation or $W^\pm$-exchange diagrams involving 
two $b$-quarks. However, due to the multiplicity of the $b$-quarks in the initial hadron, 
the decay rate is to be multiplied by the factor $N_b = 2$. 

\vspace*{3mm}
Using Eqs.~(\ref{eq:03}) and~(\ref{eq:04a}) in Eq.~(\ref{eq:08}),
for vector and axial-vector currents, we have
\begin{align}
\left\langle \mathcal{H}_{S, v^\prime}^{\left ( Q^\prime \right )} \left| 
\bar c \gamma^\mu b \right| \mathcal{H}_{A, v}^{\left ( Q \right )} \right\rangle  
& = - \xi \left ( w \right ) \text{Tr} \left\{ P_{v^\prime}^{\left ( Q \right ) \dagger} 
\frac{1 + \slashed{v}^\prime}{2} \gamma^\mu \frac{1 + \slashed{v}}{2} 
\left [ P_v^{* \left ( Q \right ) \alpha} \gamma_\alpha \gamma_5 \right ] \right\} 
\nonumber \\
& = i \xi \left ( w \right ) \varepsilon^{\mu\nu\alpha\beta} \varepsilon_\nu v_\alpha v_\beta^\prime, 
\nonumber \\
\left\langle \mathcal{H}_{S, v^\prime}^{\left ( Q^\prime \right )} \left| 
\bar c \gamma^\mu \gamma_5 b \right| \mathcal{H}_{A, v}^{\left ( Q \right )}\right\rangle  
& = - \xi \left ( w \right ) \text{Tr} \left\{ P_{v^\prime}^{\left ( Q \right ) \dagger} 
\frac{1 + \slashed{v}^\prime}{2} \gamma^\mu \gamma_5 \frac{1 + \slashed{v}}{2} 
\left[ P_v^{* \left ( Q \right ) \alpha} \gamma_\alpha \gamma_5 \right] \right\} 
\nonumber \\
& = \xi \left ( w \right ) \left[ \left ( w + 1 \right ) \varepsilon^\mu - 
\left ( \varepsilon \cdot v^\prime \right ) v^\mu \right].
\label{eq:09}
\end{align}

Comparing Eqs.~(\ref{eq:07}) and~(\ref{eq:09}), one gets
\begin{equation}
h_1 \left ( w \right ) = h_2 \left ( w \right ) = h_3 \left ( w \right ) = \xi \left ( w \right ), 
\quad 
 h_4 \left ( w \right ) = 0.  
\label{eq:10-1}
\end{equation}

Therefore, the hadronic part of the transition amplitude becomes
\begin{align}
\mathcal{M}_{T_{bb}\to T_{bc}}^{\mu} 
 & =-\frac{G_{F}}{\sqrt{2}}V_{cb}\sqrt{m_{T}m_{T^{\prime}}}\xi_{T\to T^{\prime}}\left(w\right)\left[\left(w+1\right)\varepsilon^{\mu}-\left(\varepsilon\cdot v^{\prime}\right)v^{\mu}-i\varepsilon^{\mu\nu\alpha\beta}\varepsilon_{\nu}v_{\alpha}v_{\beta}^{\prime}\right].
\label{eq:13}
\end{align}

In $B \to \left ( D,\, D^* \right ) \ell^- \nu_\ell$ decays, one has
\begin{align}
\frac{\left\langle D\left(v^{\prime}\right)\left|\bar{c}\gamma^{\mu}b\right|B\left(v\right)\right\rangle }{\sqrt{m_{B}m_{D}}} & =h_{+}\left(w\right)\left(v+v^{\prime}\right)^{\mu}+h_{-}\left(w\right)\left(v-v^{\prime}\right)^{\mu},\nonumber \\
\frac{\left\langle D^{*}\left(v^{\prime},\varepsilon^{\prime}\right)\left|\bar{c}\gamma^{\mu}b\right|B\left(v\right)\right\rangle }{\sqrt{m_{B}m_{D^{*}}}} & =ih_{V}\left(w\right)\varepsilon^{\mu\nu\alpha\beta}\varepsilon^{*\prime\nu}v^{\prime\alpha}v^{\beta},\nonumber \\
\frac{\left\langle D^{*}\left(v^{\prime},\varepsilon^{\prime}\right)\left|\bar{c}\gamma^{\mu}\gamma_{5}b\right|B\left(v\right)\right\rangle }{\sqrt{m_{B}m_{D^{*}}}} & =h_{A_{1}}\left(w\right)\left(w+1\right)\varepsilon^{*\prime\mu}-h_{A_{2}}\left(w\right) \left ( \varepsilon^{*\prime}\cdot v \right ) v^\mu - h_{A_3} \left(w\right) \left ( \varepsilon^{*\prime}\cdot v \right ) v^{\prime\mu}. 
\label{eq:10}
\end{align}
Using the procedure adopted in Eq. (\ref{eq:09}), with the field defined
in Eq. (\ref{eq:01}), one gets the well-known Isgur-Wise function for the weak transitions in $B$-meson
decays~\cite{Isgur:1989vq,Isgur:1990yhj}, i.e., 

\begin{equation}
h_+ \left ( w \right ) = h_V \left ( w \right ) = h_{A_1} \left ( w \right ) = 
h_{A_3} \left ( w \right ) = \xi^{B\to D^{\left ( * \right )}} \left ( w \right ), 
\quad 
h_- \left ( w \right ) = h_{A_2} \left ( w \right ) = 0.
\label{eq:11}
\end{equation}
In the heavy-quark symmetry limit, there is a single Isgur-Wise function, 
for both the $B \to D$ and $B \to D^*$ transitions, which is normalised at the symmetry-point: 
$\xi ^{B\to D^{\left(*\right)}}\left(w=1\right)=1 $.
Symmetry-breaking corrections of ${\cal O} (\alpha_s (m_b))$ and power corrections 
of ${\cal O} (\Lambda_{\rm QCD}/m_b)$ have been calculated~\cite{Falk:1990yz}. 
They yield $\xi^{B \to D} (w = 1) = 0.98 \pm 0.07$~\cite{Ligeti:1993hw}. 
For the $B \to D^* \ell^- \nu_\ell$, Luke's theorem protects the leading order
corrections~\cite{Luke:1990eg}, with the second-order yielding
$\xi^{B \to D^*} (w=1) = 0.91 \pm 0.03$~\cite{Neubert:1991td,Shifman:1994jh,Czarnecki:1996gu}. 
They have been used in the precise determination of the CKM matrix element $\vert V_{cb} \vert$ 
from the exclusive $B$-meson decays~\cite{Mannel:2023pdg}.

\vspace*{3mm}
Thus, in the symmetry limit, HQET relates the two form factors appearing in the semileptonic decays 
of the doubly-bottom tetraquark and $B$-mesons
\begin{equation}
\xi_{T\to T^\prime} \left ( w \right ) = 
\frac{\sqrt{m_B m_{D^{\left ( * \right )}}}}{\sqrt{m_T m_{T^\prime}}}\, 
\xi^{B\to D^{\left ( * \right )}} \left ( w \right ).
\label{eq:12}
\end{equation}
This relation will be modified including the process-dependent symmetry-breaking effects. 
They pertain to the interactions involving the spectators, which differ for the $B$-mesons 
and the doubly-heavy tetraquarks. We expect them to be subdominant and the relation 
in Eq.~(\ref{eq:12}) is a good approximation. Given a model for the doubly-heavy tetraquark 
wave-functions, they can be calculated, in principle, but are beyond the scope of this Letter.

\vspace*{3mm}
We now discuss the overall normalization related to the Fock-space composition of the tetraquarks in question. 
This is also related to the tetraquark wave-function. In this, we follow the formulation used 
in the study of the creation operators for the $T_{bb \bar u \bar d}$ tetraquark, whose ground state 
has the quantum numbers $J^P = 1^+$~\cite{Bicudo:2021qxj}, as also assumed here. The wave-function 
$\vert \Phi_{b,d} \rangle$ of this tetraquark is spanned by the two components in question, 
meson-meson, $\vert \Phi_{BB, (1 + \gamma_0)\gamma_5} \rangle$, and 
diquark-antidiquark, $\vert \Phi_{Dd, (1 + \gamma_0) \gamma_5}\rangle$.
Writing the wave-function 
\begin{equation}
\vert \Phi_{b,d} \rangle = b\,  \vert \Phi_{BB, (1 + \gamma_0)\gamma_5 \rangle}  
+ d\, \vert \Phi_{Dd, (1 + \gamma_0) \gamma_5} \rangle,
\label{eq:Bicudo1}
\end{equation}
allows to define the ratios
\begin{equation}
\omega_{BB} (r) = \frac{\vert b \vert^2}{\vert b \vert^2 + \vert d \vert^2}, 
\qquad 
\omega_{Dd} (r) = \frac{\vert d \vert^2}{\vert b \vert^2 + \vert d \vert^2},
\label{eq:Bicudo2}
\end{equation}
with $\omega_{BB} (r) + \omega_{Dd} (r) = 1$, which can be interpreted as the relative weights 
of a meson-meson and diquark-antidiquark at the $bb$-separation~$r$ in the ground state with $J^P = 1^+$. 
Lattice studies yield that the diquark-antidiquark component $\omega_{Dd} (r)$ dominates over the meson-meson 
component for $r < 0.20$~fm~\cite{Bicudo:2021qxj}. 

\vspace*{3mm}
For the general case in which both the diquarkonic and mesonic components are present in the Fock space 
of the tetraquarks, one has to project out the diquark component to determine the normalization 
of the decay rates discussed here. It is related to the fraction $\omega_{Dd} (r)$, being its integral 
obtained by integrating $\omega_{Dd} (r)$ over the size of a tetraquark. This is {\it a priori} 
not known, but the general expectations are that the compact doubly-bottom tetrahadrons should have 
a similar hadronic size as a $B$-meson. We call this quantity $f_{Dd} (bb)^2$, where we admit 
the possibility that this fraction may depend on the heavy-quark flavor, and take that into account 
as the compact-hadronic fraction of the tetraquark in numerically estimating the decay rates. 
The transition amplitudes are calculated by setting $f_{Dd} (bb)^2 = 1$ (and likewise, $f_{Dd} (bc)^2 = 1$).
The corresponding leptonic part is
\begin{equation}
\mathcal{L}_\mu = \bar{u}\left(p_1,m_{\ell}\right)\gamma_{\mu}\left(1-\gamma_5\right)v\left(p_2,m_{\nu_\ell}\right),
\label{eq:lep}
\end{equation} and the neurtino mass, $m_{\nu_\ell}$, is neglected in what follows.   
The decay width is determined by the expression
\begin{equation}
\frac{d\Gamma}{dw} = \frac{1}{(2\pi)^3}\, \frac{1}{32 m_T^3}\, 
\frac{1}{g} \left | \bar{\mathcal{A}}_{T \to T^\prime} \right |^2, 
\label{ddrate}
\end{equation}
where $\left|\bar{\mathcal{A}}_{T\to T^\prime}\right|^2$ can be calculated from the square 
of the amplitude~(\ref{eq:13}) after summing over the polarizations of initial state tetraquark, i.e., 
$\sum_\lambda \varepsilon_\mu \left ( v, \lambda \right ) \varepsilon_\nu^* \left ( v, \lambda \right ) 
= - g_{\mu\nu} + v_\mu v_\nu$, contracting it with the square of the spin-summed leptonic amplitude~(\ref{eq:lep}), and integrating over everything except the HQET parameter $w$. 
The factor~$1/g$ above denotes the spin average of the initial state tetraquark. 
For $J^P = 1^+$, the number of the tetraquark spin polarizations is $g = 3$.

Using this, along with the decay kinematics in Eq.~(\ref{ddrate}), the differential decay width is expressed as
\begin{eqnarray}
\frac{d\Gamma}{dw} &=& \frac{N_b}{g}\, \frac{G_F^2 \left|V_{cb}\right|^2 m_{T^\prime}}{384\pi^3 m_{T}^2}\left|\xi_{T\to T^\prime}\left(w\right)\right|^2\sqrt{w^2-1}\sqrt{1-\frac{m_{\ell}^2}{s}}\bigg[3m_{\ell}^4\left(\left(m_{T}+m_{T^\prime}\right)^2+4m_{T}m_{T^\prime}w\right)
     \notag\\
     &&-\frac{4m_{\ell}^2}{s}m_{T}m_{T^\prime}\left(w+1\right)\left(6m_{T}^4+m_{T}^3m_{T^\prime}\left(5-11w\right)-2m_{T}^2m_{T^\prime}^2\left(2w^2+5w-1\right)+m_{T}m_{T^\prime}^3\left(13w+5\right)-6m_{T^\prime}^4\right)\notag\\
&&+8m_{T}^2 m_{T^\prime}^2\left(w+1\right)\left(\left(m_{T}^2+m_{T^\prime}^2\right)\left(5w+1\right)-2m_{T}m_{T^\prime}\left(4w^2+w+1\right)\right)\bigg],\label{drate-mell}
\end{eqnarray}
where $s = q^2 = m_T^2 + m_{T^\prime}^2 - 2 m_T m_{T^\prime} w$ is the momentum transfer squared, 
and $N_b = 2$ for the $T_{bb \bar u \bar d}$ state. 

In the massless charged-lepton case ($m_\ell = 0$), Eq.~(\ref{drate-mell}) simplifies to
\begin{equation}
\frac{d\Gamma}{dw} = N_b\, \frac{G_F^2}{144\pi^3} \left | V_{cb} \right |^2 m_{T^\prime}^3 \left | \xi_{T \to T^\prime} 
\left ( w \right ) \right |^2 \left (w + 1 \right )^{3/2} \sqrt{w - 1} \left [ 
\left ( m_T^2 + m_{T^\prime}^2 \right ) \left ( 5w + 1 \right ) 
- 2 m_T m_{T^\prime} \left ( 4 w^2 + w + 1 \right ) \right ].
\label{eq:14}
\end{equation}

Using the form factors relation defined in Eqs.~(\ref{eq:12}) and~(\ref{eq:14}) yields
\begin{align}
\frac{d\Gamma}{dw} & = N_b\, \frac{G_F^2}{144\pi^{3}} \left | V_{cb} \right |^2 
\frac{m_B m_D}{m_T}\, m_{T^\prime}^2 \left | \xi^{B\to D} \left ( w \right ) \right |^2
\left ( w + 1 \right)^{3/2} \sqrt{w - 1} \left ( f_{Dd} (bb) \right )^2 
\nonumber \\
& \times \left[\left(m_T^2 + m_{T^\prime}^2 \right)\left(5w+1\right)-2 m_T m_{T^\prime}\left(4w^2+w+1\right)\right], 
\label{eq:14b}
\end{align}
where the additional factor $\left ( f_{Dd} (bb) \right )^2$ indicates 
that the tetraquark is decaying through its diquark-antidiquark component.
To calculate the lepton energy spectrum, and the semileptonic decay rate, we need 
to parametrize the Isgur-Wise function for the weak decays of the heavy-to-heavy tetraquarks. 
We use the corresponding form for the $B \to (D, D^*)$ form factors in the zero recoil expansion 
(c.f. Eq.~(41) in~\cite{Caprini:1997mu}):
\begin{equation}
\xi^{B\to D}\left(w\right) = \xi^{B\to D}\left(1\right) \left[ 1 - 8 \rho_1^2 z + \left ( 51 \rho_1^2 - 10 \right ) z^2 - \left ( 252 \rho_1^2 - 84 \right ) z^3 \right],
\label{eq:16}
\end{equation}
where $z = \left ( \sqrt{w + 1} - \sqrt 2 \right )/\left ( \sqrt{w + 1} + \sqrt 2 \right )$, 
and $\rho_1^2$ is a slope parameter at zero-recoil, bounded between $-0.14 < \rho_1^2 < 1.54$. 
We anticipate that the slope of the Isgur-Wise functions in the two cases, namely $B \to (D, D^*)$ 
and $T_{bb \bar u \bar d} \to T_{bc \bar u \bar d}$, are expected to differ from each other, 
but for the sake of definiteness, we shall use the same range for the two cases. 
The rationale behind this assumption is that in both cases, the recoil momenta 
of the spectators are similar and small.


\vspace*{3mm}
Using the numerical values for the various input parameters given in Table~\ref{tab:MesonPar}, and 
with $\left ( f_{Dd} (bb) \right )^2 = 0.5$, as indicated by the Lattice-QCD studies~\cite{Bicudo:2021qxj},
yields the following decay width:
\begin{align}
\Gamma \left (T_{bb \bar u \bar d} (J^P = 1^+) \to T_{bc \bar u \bar d} (J^P = 0^+)\, \ell^-\nu_\ell \right ) 
& = \left ( 0.32,\; 0.11 \right )\; \frac{\left ( f_{Dd} (bb)\right )^2}{0.5} \times 10^{-11}\, \text{MeV}.
\label{eq:16b}
\end{align}
The first number corresponds to the $e^- \nu_e$ and $\mu^- \nu_\mu$ cases, and the second for 
the $\tau^-\nu_\tau$ case. Further studies are needed to quantify $\left ( f_{Dd} (bb) \right )^2$. 
However, this can also be determined by the branching ratios discussed here, once they are measured. 

Using the total width $\Gamma_{\rm total} (T_{bb \bar u \bar d}) = 8.2 \times 10^{-10}$~MeV, 
which is derived from the lifetime $\tau (T_{bb \bar u \bar d}) = 0.8$~ps~\cite{Ali:2018ifm}, 
we get the following branching ratios:
\begin{align}
{\cal B} \left(T_{bb\bar{u}\bar{d}} (J^P=1^+) \to T_{bc\bar{u}\bar{d}} (J^P=0^+)\, \ell^{-}\nu_{\ell}\right) & = 0.4\%~({\text{for}}~ \ell = e,\; \mu), 
\nonumber \\ 
{\cal B} \left(T_{bb\bar{u}\bar{d}} (J^P=1^+) \to T_{bc\bar{u}\bar{d}} (J^P=0^+)\, \tau^{-}\nu_{\tau}\right) & = 1.4 \times 10^{-3}.
\end{align}


\section{
Semileptonic decays $T_{bb \bar u \bar d} \left ( J^P = 1^+ \right ) 
\to T_{bc \bar u \bar d} \left ( J^P = 1^+ \right ) \ell^- \nu_\ell$} 
\label{sec:3}

In this section, we will consider the final state tetraquark $T_{bc \bar u \bar d}$ with $J^P = 1^+$, 
therefore, $T_{bb \bar u \bar d} \to T_{bc \bar u \bar d}$ corresponds to the axial-vector to axial-vector 
transition. With the effective Hamiltonian describing the $b\to c \ell^- \nu_\ell$ transition given 
in Eq.~(\ref{eq:06}), the matrix elements in HQET can be parameterized as
\begin{align}
\frac{\left\langle T_{bc \bar u \bar d} \left (v^\prime, \varepsilon^\prime \right ) \left | 
      \bar c \gamma^\mu b \right | T_{bb \bar u \bar d} \left ( v, \varepsilon \right ) \right\rangle}
     {\sqrt{m_T m_{T^\prime}}} 
& = - \left( \varepsilon^{\prime *} \cdot \varepsilon \right ) \left [ 
h_1 \left ( w \right ) \left ( v + v^\prime \right )^\mu + 
h_2 \left ( w \right ) \left ( v - v^\prime \right )^\mu 
\right ] 
\nonumber \\
& + h_3 \left ( w \right ) \left ( \varepsilon^{\prime *} \cdot v \right ) \varepsilon^\mu 
  + h_4 \left ( w \right ) \left ( \varepsilon \cdot v^\prime \right ) \varepsilon^{\prime * \mu} 
  - \left ( \varepsilon \cdot v^\prime \right ) \left ( \varepsilon^{\prime *} \cdot v \right ) 
    \left [ h_5 \left ( w \right ) v^\mu + h_6 \left ( w \right ) v^{\prime\mu} \right ], 
\nonumber \\
\frac{\left\langle T_{bc \bar u \bar d} \left ( v^\prime, \varepsilon^\prime \right ) \left | 
      \bar c \gamma^\mu \gamma_5 b \right | T_{bb \bar u \bar d} \left (v, \varepsilon \right ) \right\rangle}
     {\sqrt{m_T m_{T^\prime}}} & = i \varepsilon^{\mu\nu\alpha\beta} 
\left \{ \varepsilon_\alpha \varepsilon_\beta^{\prime *} \left [ 
h_7 \left ( w \right ) \left ( v + v^\prime \right )_\nu + 
h_8 \left ( w \right ) \left ( v - v^\prime \right )_\nu \right ] \right. 
\nonumber \\
& + v_\alpha^\prime v_\beta \left [ 
h_9 \left ( w \right ) \left ( \varepsilon^{\prime *} \cdot v \right ) \varepsilon_\nu + 
h_{10} \left ( w \right ) \left ( \varepsilon \cdot v^\prime \right ) \varepsilon_\nu^{\prime *} \right],
\label{eq:17}
\end{align}
where $m_T$ and $m_{T^\prime}$ denote the masses of the initial and final state tetraquarks, 
$T_{bb \bar u \bar d}$ and $T_{bc \bar u \bar d}$, respectively. 

For the $1^+ \to 1^+$ transitions, the heavy-quark spin symmetry implies
\begin{equation}
\left\langle \mathcal{H}_{A, v^\prime}^{\left ( Q^\prime \right )} \left | 
\bar Q^\prime \Gamma Q \right | \mathcal{H}_{A, v}^{\left ( Q \right )} \right\rangle 
= - \xi \left ( w \right ) \text{Tr} \left\{ \bar{\mathcal{H}}_{A, v^\prime}^{\left ( Q^\prime \right )} 
\Gamma \mathcal{H}_{A, v}^{\left ( Q \right )} \right \}, 
\label{eq:18}
\end{equation}
which on using Eqs.~(\ref{eq:03}) and~(\ref{eq:04}) gives
\begin{align}
\left\langle \mathcal{H}_{A, v^\prime}^{\left ( Q^\prime \right )} \left | 
\bar c \gamma^\mu b \right | \mathcal{H}_{A, v}^{\left ( Q \right )} \right\rangle  
& = - \xi \left ( w \right ) \text{Tr} \left\{ 
\left [ P_{v^\prime}^{* \left ( Q^\prime \right ) \dagger \nu} \gamma_\nu \gamma_5 \right ] 
\frac{1 + \slashed{v}}{2} \gamma^\mu \frac{1 + \slashed{v}}{2} 
\left [ P_v^{* \left ( Q \right ) \alpha} \gamma_\alpha \gamma_5 \right ] 
\right\} 
\nonumber \\
& = \xi \left ( w \right ) \left\{ 
\left ( \varepsilon^{\prime *} \cdot \varepsilon \right ) \left ( v + v^\prime \right )^\mu - 
\left ( \varepsilon^{\prime *} \cdot v \right ) \varepsilon^\mu - 
\left ( \varepsilon \cdot v^\prime \right ) \varepsilon^{\prime * \mu} \right\} 
\nonumber \\
\left\langle \mathcal{H}_{A, v^\prime}^{\left ( Q^\prime \right )} \left | 
\bar c \gamma^\mu \gamma_5 b_v \right | \mathcal{H}_{A, v}^{\left ( Q \right )} \right\rangle  
& = - \xi \left ( w \right ) \text{Tr} \left\{ 
\left [ P_{v^\prime}^{* \left ( Q^\prime \right ) \dagger \nu} \gamma_\nu \gamma_5 \right ] 
\frac{1 + \slashed{v}}{2} \gamma^\mu \gamma_5 \frac{1 + \slashed{v}}{2} 
\left [ P_v^{* \left ( Q \right ) \alpha} \gamma_\alpha \gamma_5 \right ] 
\right\}  
\nonumber \\
& = i \xi \left ( w \right ) \varepsilon^{\mu\nu\alpha\beta} 
\varepsilon_\alpha \varepsilon_\beta^{\prime *} \left ( v + v^\prime \right )_\nu . 
\label{eq:19}
\end{align}
Here, we have written $P_{v^\prime}^{* \left ( Q^\prime \right ) \dagger \nu}$
and $P_v^{* \left ( Q \right ) \alpha}$ as the polarization vectors $\varepsilon^{\prime * \nu}$
and $\varepsilon^\alpha$, respectively, and used 
$\left ( \varepsilon^{\prime *} \cdot v^\prime \right ) = \left ( \varepsilon \cdot v \right ) = 0$. 

Comparing Eq.~(\ref{eq:17}) and~(\ref{eq:19}), we get
\begin{equation}
h_1 \left ( w \right ) = h_3 \left ( w \right ) = h_4 \left ( w \right ) = 
h_7 \left ( w \right ) = \xi \left ( w \right ),
\label{eq:19-rel}
\end{equation}
and the others vanish in this limit. Hence, the relation given in Eq.~(\ref{eq:12}) holds in this case too. 

The corresponding hadronic part of the decay amplitude can be written as
\begin{align}
\mathcal{M}^\mu_{T_{bb \bar u \bar d} \to T_{bc \bar u \bar d}^\prime} & = \frac{G_F}{\sqrt 2}\, V_{cb} 
\left\langle T_{bc \bar u \bar d} \left (v^\prime, \varepsilon^\prime \right ) \left | 
\bar c \gamma^\mu \left ( 1 - \gamma_5 \right ) b_v 
\right | T_{bb \bar u \bar d} \left ( v, \varepsilon \right ) \right\rangle 
\nonumber \\
& = - \frac{G_F}{\sqrt 2}\, V_{cb}\, \xi_{T \to T^\prime} \left ( w \right ) \left [ 
\left ( \varepsilon^{\prime *} \cdot \varepsilon \right ) \left ( v + v^\prime \right )^\mu - 
\left ( \varepsilon^{\prime *} \cdot v \right ) \varepsilon^\mu - 
\left ( \varepsilon \cdot v^\prime \right ) \varepsilon^{\prime * \mu} + 
i \varepsilon^{\mu\nu\alpha\beta} \varepsilon_\alpha \varepsilon_\beta^{\prime *} 
\left ( v + v^\prime \right )_\nu \right ].
\label{eq:ax-12a}
\end{align}
Contracting the above hadronic amplitude with the leptonic part~(\ref{eq:lep}) 
and summing over the tetraquark polarizations, i.e., $\sum_\lambda \varepsilon_\mu 
\left ( v^{(\prime)},\lambda \right ) \varepsilon_\nu^* \left (v^{(\prime)},\lambda \right ) 
= - g_{\mu\nu} + v^{(\prime)}_\mu v^{(\prime)}_\nu$, and spins of the leptons, 
the differential decay  width becomes
\begin{eqnarray}
\frac{d\Gamma^{1^+\to 1^+}}{dw}&=&\frac{N_b}{g}\, 
\frac{G_F^2 \left | V_{c b} \right |^2 m_{T^\prime}}{384\pi^3 m_{T}^2}\left|\xi_{T\to T^\prime}\left(w\right)\right|^2\sqrt{w^2-1}\sqrt{1-\frac{m_{\ell}^2}{s}}\bigg[-3m_{\ell}^4\left(m_{T}^2+m_{T^\prime}^2+2m_{T}m_{T^\prime}\left(4w+5\right)\right)\notag\\
&&-\frac{4m_{\ell}^2}{s}m_{T}m_{T^\prime}\left(w+1\right)\left(12m_{T}^4+m_{T}m_{T^\prime}\left(m_{T}^2+m_{T^\prime}^2\right)\left(1-49w\right)+2m_{T}^2m_{T^\prime}^2\left(20w^2-w+17\right)+12m_{T^\prime}^4\right)\notag\\
&&+8m_{T}^2 m_{T^\prime}^2\left(w+1\right)\left(\left(m_{T}^2+m_{T^\prime}^2\right)\left(13w-1\right)-2m_{T}m_{T^\prime}\left(8w^2-w+5\right)\right)\bigg],
\label{drate-axial-lm}
\end{eqnarray}
which in the massless charged-lepton case reduces to
\begin{eqnarray}
\frac{d\Gamma^{1^+\to 1^+}}{dw}&=& N_b\, \frac{G_F^2 \left | V_{cb}\right |^2}{144\pi^3} \left|\xi_{T\to T^\prime}\left(w\right)\right|^2 m^3_{T^\prime}(1+w)^{3/2}\sqrt{w-1} 
\notag\\
&& \times \left[\left(m^2_{T^\prime}+m^2_{T}\right)\left(13w-1\right)-2m_{T}m_{T^\prime}\left(8w^2-w+5\right)\right] .\label{drate-axial}
\end{eqnarray}
Using the numerical values of the input parameters from Table \ref{tab:MesonPar}, and integrating over $w$ in the allowed kinematic range gives
\begin{eqnarray}
\Gamma\left(T_{bb\bar{u}\bar{d}}\left(J^P=1^+\right)\to T_{bc\bar{u}\bar{d}}\left(J^P=1^+\right)\ell^{-}\nu_{\ell}\right) & = & 1.92\;\frac{\left(f_{Dd}(bb)\right)^{2}} {0.5}\times10^{-11}\;  \text{MeV};\; ({\text{for}}~ \ell = e,\; \mu) 
\nonumber \\
\Gamma\left(T_{bb\bar{u}\bar{d}}\left(J^P=1^+\right)\to T_{bc\bar{u}\bar{d}}\left(J^P=1^+\right)\tau^{-}\nu_{\tau}\right) & = & 0.88\;
\frac{\left(f_{Dd}(bb)\right)^{2}} {0.5}\times10^{-11}\;  \text{MeV} .
\label{rate-1p1p}  
\end{eqnarray}
They yield the following branching ratios:
\begin{align}
 {\cal B}(T_{bb\bar{u}\bar{d}} (J^P=1^+) \to T_{bc\bar{u}\bar{d}} (J^P=1^+)\ell^{-}\nu_{\ell}) & = 1.2\%~({\text{for}}~ \ell = e,\; \mu),
\nonumber \\ 
 {\cal B} (T_{bb\bar{u}\bar{d}} (J^P=1^+) \to T_{bc\bar{u}\bar{d}} (J^P=1^+)\tau^{-}\nu_{\tau}) & = 0.53 \%.
\end{align}
%
Thus,  ${\mathcal{B}} \left(T_{bb\bar{u}\bar{d}}\left(J^P=1^+\right)\to T_{bc\bar{u}\bar{d}}\left(J^P=0^+\right)\ell^-\nu_\ell\right)$ and 
$\mathcal{B}\left(T_{bb\bar{u}\bar{d}}\left(J^P=1^+\right)\to T_{bc\bar{u}\bar{d}}\left(J^P=1^+\right)
\ell^- \nu_\ell\right)$ 
are together about 1.6\%.
We expect similar branching ratios for the decays 
$T_{bb\bar u \bar d} \to \Xi_{bc}^+ \;\bar{p} \;\ell^- \nu_\ell$ and $T_{bb\bar u \bar d} \to \Xi_{bc}^0 \;\bar{n} \; \ell^- \nu_\ell$. 
\begin{table}[tbp]
\renewcommand{\arraystretch}{1.7}
\begin{tabular}{|c|}
\hline\hline
$m_{T_{bb \bar u \bar d}}\left(1^+\right) = 10,504.4^{+37.36}_{-44.15}~\mathrm{MeV}$, \quad
$m_{T_{bc \bar u \bar d}}\left(0^+\right) = 7155^{+9}_{-19}~\mathrm{MeV}$, \quad
$m_{T_{bc \bar u \bar d}}\left(1^+\right) = 7152^{+9}_{-19}~\mathrm{MeV}$, \quad
$m_{T_{cc \bar u \bar d}}\left(1^+\right) = 3875~\mathrm{MeV}$ \\
$m_\pi = \left(139.57039\pm 0.00018\right)~\mathrm{MeV}$, \quad
$m_\rho = \left(775.26\pm 0.23\right)~\mathrm{MeV}$, \quad
$m_{a_1} = \left(1230\pm 40\right)~\mathrm{MeV}$, \quad
$m_D =(1869.61 \pm 0.10)~\mathrm{MeV}$ \\
$m_B =(5279.66\pm 0.12)~\mathrm{MeV}$, \quad
$m_{B^*}=(5324.71\pm 0.21)~\mathrm{MeV}$, \quad
$f_\pi =\left(130.41\pm 0.03\pm 0.20\right)~\mathrm{MeV}$, \quad
$f_\rho =\left(210\pm 4\right)~\mathrm{MeV}$  \\
$f_{a_1 }=234~\mathrm{MeV}$, \quad
$\rho_1^2 = 1.3$, \quad
$C_1\left(m_b\right)\left[C_2\left(m_b\right)\right]=1.117\left[-0.257\right]$, \quad
$G_F = 1.16637\times10^{-11}~\mathrm{MeV}^{-2}$,\quad 
$m_e = 0.511~\mathrm{MeV}$\\
$m_\mu = 105.7~\mathrm{MeV}$,\quad 
$m_\tau = 1777~\mathrm{MeV}$,\quad 
$V_{ud}=0.97420\pm 0.3021$, \quad 
$|V_{cb}| = \left(39.5\pm 0.08\right)\times10^{-3}$,\quad 
$V_{cs} = \left(0.997\pm 0.017\right)$\\
$f_{D^-_s} = \left(257.8\pm 4.1\right)~\mathrm{MeV}$, \quad 
$m_{D^-_s} = \left(1968.3\pm 0.11\right)~\mathrm{MeV}$, \quad  
$f_{D^{*-}_s} = 315~\mathrm{MeV}$ , $m_{D^{*-}_s} = \left(2112.2\pm 0.4\right)~\mathrm{MeV}$\\
\hline\hline
\end{tabular}%
\caption{Numerical values of the input parameters. 
Decay rates are calculated for the central values in this table.}
\label{tab:MesonPar}
\end{table}
\section{Non-Leptonic Decays $T_{bb \bar u \bar d} \to T_{bc \bar u \bar d}\; h^-$}
\label{sec:4}
The simplest non-leptonic decay of the doubly-bottom tetraquark is 
$T_{bb \bar u \bar d} \to T_{bc \bar u \bar d} h^-$, where $h^-$ is a light meson, 
such as $h^- =\pi^-, \rho^-, a_1^-$, or the heavy-light meson $D^-_s$ and $D^{*-}_s$. 
At the quark level, these decays take place via the weak transition 
$b \to c W^-\, (\to \bar u d,\; \bar c s)$. This is similar to the $B$-meson decays, 
such as $B \to D^{(*)} \left ( \pi,\; D_s \right )$, which were calculated in the so-called 
``naive factorization approach''~\cite{Bauer:1986bm}. Subsequent improvements have provided
a QCD basis for this class of factorized amplitudes, and the resulting corrections are found 
to be small~\cite{Dugan:1990de}. We assume that this factorization approach can also be used 
to study the non-leptonic decays of the tetraquarks, though a formal proof is lacking.

\begin{figure}[!htb]
\centering\scalebox{1}{
\begin{tabular}{cccc}
\includegraphics[width=2.5in,height=1.8in]{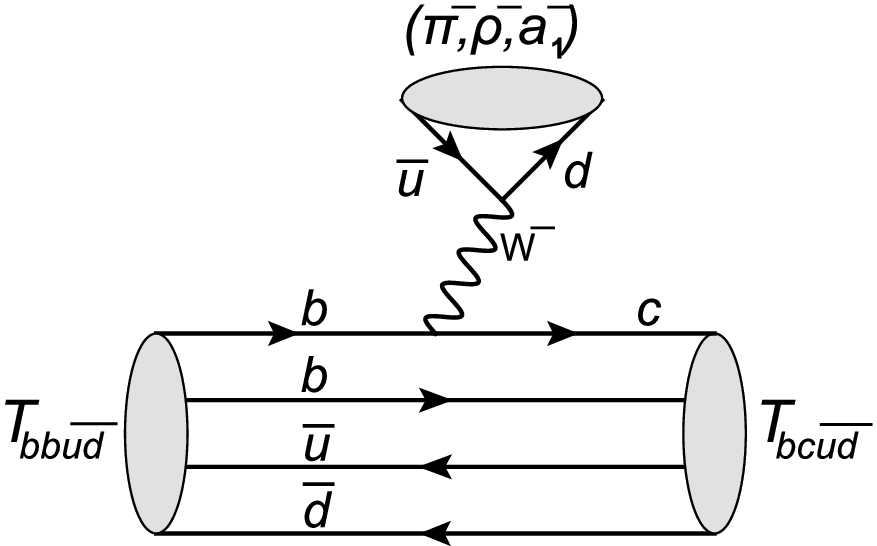}\hspace{0.5cm}\qquad & \qquad
\includegraphics[width=2.5in,height=1.8in]{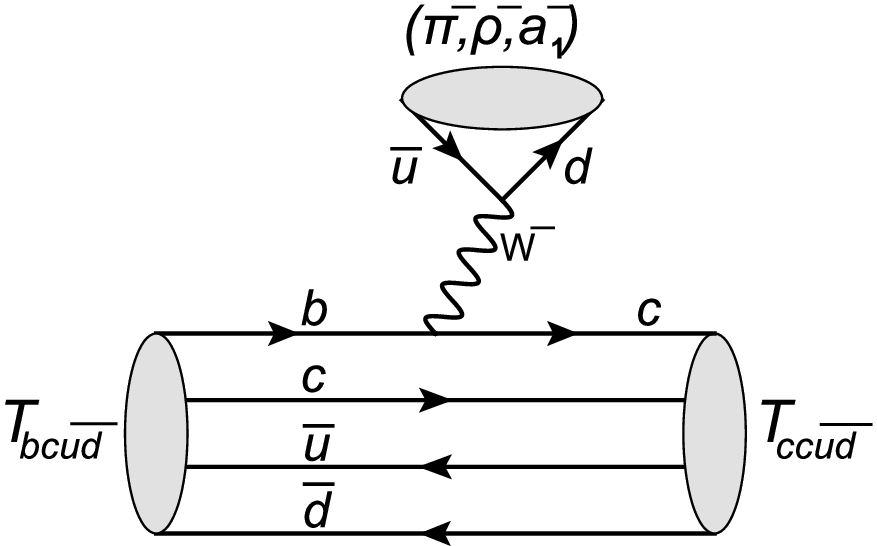} \\
(a)\hspace{0.5cm}\qquad &  \qquad (b)\\
\\
\includegraphics[width=2.5in,height=1.8in]{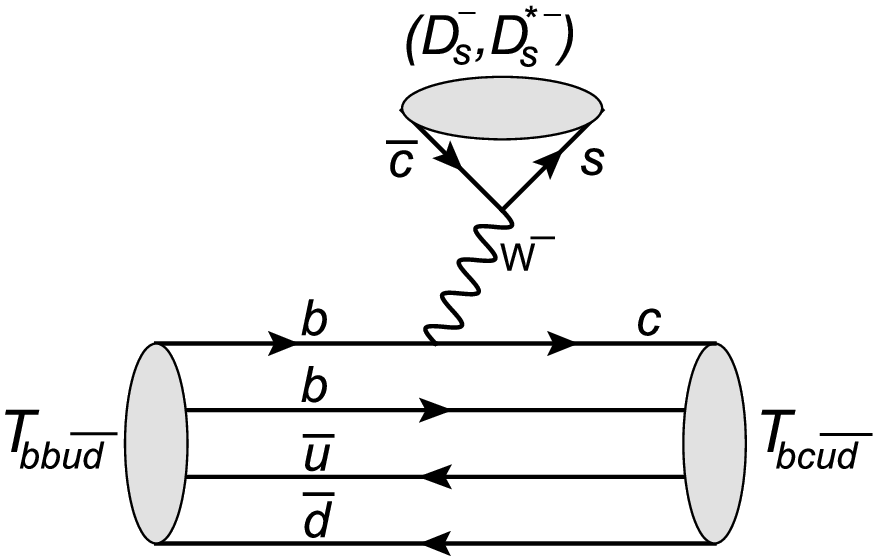}\hspace{0.5cm}\qquad & \qquad
\includegraphics[width=2.5in,height=1.8in]{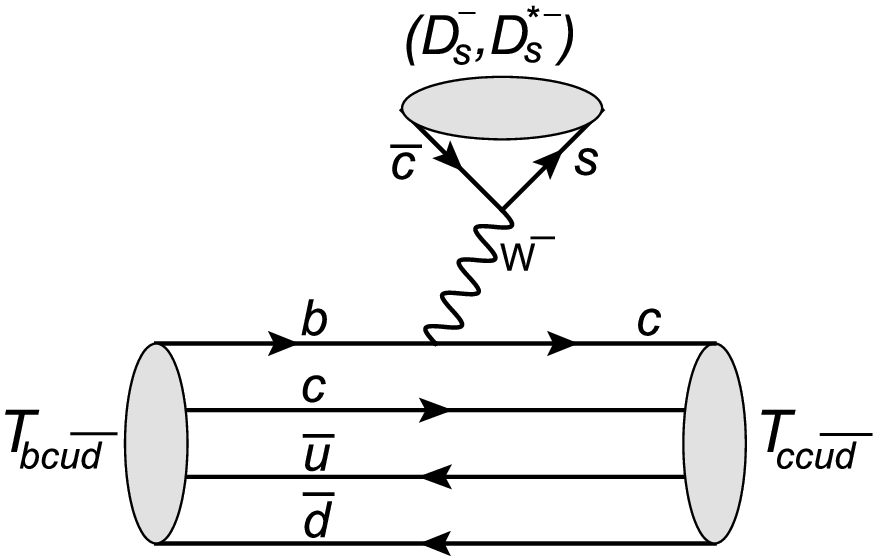} \\
\vspace{10mm}
(c)\hspace{0.5cm}\qquad &  \qquad (d)
\end{tabular}}
\caption{Feynman diagrams for 
(a) $T_{bb \bar u \bar d} \to T_{bc \bar u \bar d}\; \left ( \pi^-,\, \rho^-,\, a_1^- \right )$, 
(b) $T_{bc \bar u \bar d} \to T_{cc \bar u \bar d}\; \left ( \pi^-,\, \rho^-,\, a_1^- \right )$, 
(c) $T_{bb \bar u \bar d} \to T_{bc \bar u \bar d}\; \left ( D^-_s,\, D^{*-}_s \right )$, and 
(d) $T_{bc \bar u \bar d} \to T_{cc \bar u \bar d}\; \left ( D^-_s,\, D^{*-}_s \right )$.
}
\label{FeynDiag2}
\end{figure}

\vspace*{3mm}
The effective Hamiltonian for these decays at the tree level is given by
\begin{equation}
\mathcal{H}_{\text{eff}} = 4\, \frac{G_F}{\sqrt 2} \, V_{cb} V_{qq^\prime}^* 
\left [ C_1 \left ( \mu \right ) \mathcal{O}_1 + C_2 \left ( \mu \right ) \mathcal{O}_2 \right ], 
\label{eq:20}
\end{equation}
where $C_{1,2} \left ( \mu \right )$ are the Wilson coefficients, calculated at the factorization scale~$\mu$. 
The $\left ( q q^\prime \right )$ represents $\left ( \bar u d \right )$ and $\left ( \bar c s \right )$ 
for $\pi^-,\; \rho^-,\; a_1^-$, and $D_s,\; D^{*-}_s$, respectively. The four-quark operators are:
\begin{align}
\mathcal{O}_1 & = \left ( \bar q^\prime_i \gamma_\mu P_L q_i \right ) \left ( \bar c_j \gamma^\mu P_L b_j \right ), 
\nonumber \\
\mathcal{O}_2 & = \left ( \bar q^\prime_i \gamma_\mu P_L q_j \right ) \left ( \bar c_j \gamma^\mu P_L b_i \right ), 
\label{eq:21}
\end{align}
with $i,\, j$ representing the color indices. The current in each bracket for $\mathcal{O}_1$ 
and $\mathcal{O}_2$ is the color singlet and octet, respectively. 
The corresponding Feynman diagrams are shown in Fig.~\ref{FeynDiag2}(a) and\ref{FeynDiag2}(c). 
For the color-allowed  tree amplitudes, factorization
of the decay amplitude is expected to be a good approximation, and the decay amplitude 
for $T_{bb \bar u \bar d} \left ( J^P = 1^+ \right ) \to T_{bc \bar u \bar d} \left ( J^P = 0^+,\; 1^+ \right ) \pi^-$ 
can be written in the form:
\begin{align}
\mathcal{M} & = 4\, \frac{G_F}{\sqrt 2}\, V_{cb} V_{ud}^*\, a_1 \left ( \mu \right ) 
\left\langle \pi^- \left ( q \right ) \left | \bar d_i \gamma_\mu P_L u_i \right | 0 \right\rangle 
\left\langle T_{bc \bar u \bar d} \left ( p^\prime \right ) 
\left | \bar c_j \gamma^\mu P_L b_j \right | T_{bb \bar u \bar d} \left ( p \right ) \right\rangle 
\nonumber \\
& = -2\, \frac{G_F}{\sqrt 2}\, V_{cb} V_{ud}^*\, a_1 \left ( \mu \right ) 
\left\langle \pi^- \left ( q \right ) \left | \bar d_i \gamma_\mu \gamma_5 u_i \right | 0 \right\rangle 
\left\langle T_{bc \bar u \bar d} \left ( p^\prime \right ) 
\left | \bar c_j \gamma^\mu P_L b_j \right | T_{bb \bar u \bar d} \left ( p \right ) \right\rangle 
\nonumber \\
& = -2i\, \frac{G_F}{\sqrt 2}\, V_{cb} V_{ud}^*\, a_1 \left ( \mu \right ) f_\pi q_\mu 
\left\langle T_{bc \bar u \bar d} \left ( p^\prime \right ) 
\left | \bar c_j \gamma^\mu P_L b_j \right | T_{bb \bar u \bar d} \left ( p \right ) \right\rangle, 
\label{eq:22}
\end{align}
where, we have used $\left\langle \pi \left ( q \right ) \left | \bar d_i \gamma_\mu\gamma_5 u_i \right | 0 \right\rangle = i f_\pi q_\mu$, with $f_\pi$ representing the decay constant of $\pi^-$, $q = p - p^\prime$, and 
$a_1 \left ( \mu \right ) = C_1 \left ( \mu \right ) + C_2 \left ( \mu \right )/3$. 
Using the result of the matrix element from Eq.~(\ref{eq:13}), which holds in the heavy-quark 
symmetry limit, and noting that the tetraquark in the final state has $J^P = 0^+$, the HQET 
version of the amplitude given in Eq.~(\ref{eq:22}) takes the form
\begin{equation}
\mathcal{M}^{1^+ \to 0^+} = -i\, \frac{G_F}{\sqrt 2}\, V_{cb} V_{ud}^*\, a_1 \left ( \mu \right ) 
f_\pi\, \xi_{T \to T^\prime} \left ( w \right ) q_\mu \left [ 
\left ( w + 1 \right ) \varepsilon^\mu - \left ( \varepsilon \cdot v^\prime \right ) v^\mu 
- i \varepsilon^{\mu\nu\alpha\beta} \varepsilon_\nu v_\alpha v_\beta^\prime \right ] ,  
\label{eq:23}
\end{equation}

Along the same lines, for the final state $T_{bc \bar u \bar d}$ having $J^P = 1^+$, we use 
Eq.~(\ref{eq:ax-12a}) in Eq.~(\ref{eq:22}), which gives
\begin{equation}
\mathcal{M}^{1^+ \to 1^+} = i\, \frac{G_F}{\sqrt 2}\, V_{cb} V_{ud}^*\, a_1 \left ( \mu \right ) 
f_\pi\, \xi_{T \to T^\prime} \left ( w \right ) q_\mu \left [ 
\left ( \varepsilon^{\prime *} \cdot \varepsilon \right ) \left ( v + v^\prime \right )^\mu - 
\left ( \varepsilon^{\prime *} \cdot v \right ) \varepsilon^\mu - 
\left ( \varepsilon \cdot v^\prime \right ) \varepsilon^{\prime * \mu} + 
i\varepsilon^{\mu\nu\alpha\beta} \varepsilon_\alpha \varepsilon_\beta^{\prime *} 
\left ( v + v^\prime \right )_\nu \right ]. 
\label{eq:24}
\end{equation}

For $h = \rho$ and $h = a_1$, which are light vector and axial-vector mesons, respectively, 
the relevant matrix elements are
\begin{eqnarray}
\left\langle \rho \left ( q \right ) \left | \bar d_i \gamma_\mu u_i \right | 0 \right\rangle & = & f_\rho\, m_\rho\, \varepsilon^*_{1 \mu} \left ( q \right),
\notag\\
\left\langle a_1 \left ( q \right ) \left | \bar d_i \gamma_\mu \gamma_5 u_i \right | 0 \right\rangle & = & f_{a_1}\, m_{a_1} \varepsilon^*_{1 \mu} \left ( q \right ),
\label{eq.25}
\end{eqnarray}
where $f_\rho$ and $f_{a_1}$ are their respective decay constants and 
$\varepsilon^*_{1 \mu} \left ( q \right )$ denote their polarization vectors. 
The corresponding amplitudes involving spin-1 particles in the final state can be obtained 
by replaxcing $f_\pi$ with $f_\rho$ and $f_{a_1}$ for the $\rho$- and $a_1$-mesons 
in Eqs.~(\ref{eq:23}) and (\ref{eq:24}), respectively. Also, the momentum vector~$q_\mu$ is to be 
replaced with their mass times the polarization vector $\varepsilon^*_{1 \mu} \left ( q \right )$.

The decay width of a two-body process has the standard form~\cite{ParticleDataGroup:2022pth}:
\begin{equation}
\Gamma = \frac{1}{8\pi}\, \frac{1}{g} \left | \mathcal{\bar{M}} \right|^2 \frac{\left|\bf{p}_1\right|}{M^2}, 
\label{2body-rate}
\end{equation}
where $g = 3$ in this case, and the three-momenta of the final state particles in the rest frame 
of the parent hadron given by
\begin{equation}
\left | {\bf p}_1 \right | = \left | {\bf p}_2 \right | = 
\frac{1}{2M}\, \sqrt{\left ( M^2 - \left ( m_1 + m_2 \right)^2 \right ) \left ( M^2 - \left ( m_1 - m_2 \right )^2 \right)} 
 \equiv \frac{1}{2M}\, \lambda \left ( M^2, m_1^2, m_2^2 \right ) .
\label{three-momta}
\end{equation}
Here, $m_1$ and $m_2$ are the masses of final state particles and $M$ is the mass of the decaying particle. 
Using Eq.~(\ref{eq:23}), we get the  width of the  
$T_{bb \bar u \bar d} \to T_{bc \bar u \bar d}\, \pi^-$ decay:  
\begin{equation}
\Gamma^{1^+ \to 0^+} = \frac{N_b}{g}\, 
\frac{G_F^2 \left|V_{cb}V_{ud}^*\right|^2 a_1^2 (\mu) f_\pi^2}{512\pi m_T^6 m_{T^\prime}} 
\left | \xi_{T \to T^\prime} \left ( w \right ) \right |^2 
\lambda^{3/2} \left (m_T^2, m_{T^\prime}^2, m_{\pi^-}^2 \right ) 
\left [ m_T^2 + 2 m_T m_{T^\prime} \left ( w + 1 \right ) - m_{T^\prime}^2 + m_{\pi^-}^2 \right ]^2, 
\label{drate-1zpi} 
\end{equation}
where 
$w$ is evaluated at the final state hadron $\left ( \pi^- \right )$ mass, i.e., 
\begin{equation}
w = \frac{m_T^2 + m_{T^\prime}^2 - m_{\pi^-}^2}{2 m_T m_{T^\prime}}.
\label{w-def}
\end{equation}
For the $1^+ \to 1^+$ decay, we use Eq.~(\ref{eq:24}) to get
\begin{eqnarray}
 \Gamma^{1^+\to 1^+} &=& N_b\, \frac{G_F^2\left|V_{cb}V_{ud}^*\right|^2 a_1^2 (\mu) f_\pi^2}{384\pi m_{T}^{4}m_{T^\prime}}\left|\xi_{T\to T^\prime}\left(w\right)\right|^2 \\
 &&\times \lambda^{1/2} \left(m_{T}^2,m_{T^\prime}^2,m_{\pi^-}^2\right) \left(\left(m_{T}+m_{T^\prime}\right)^2-m_{\pi^-}^2\right)\left(m_{\pi^-}^2\left(2m_T m_{T^\prime}-5\left(m_{T}^2+m_{T^\prime}^2\right)\right)+5\left(m_{T}^2-m_{T^\prime}^2\right)^2\right).\notag
 \label{drate-11pi} 
\end{eqnarray}
Using the values of decay constant $f_\pi$ and the other input parameters from Table~\ref{tab:MesonPar},
setting $N_b=2$ and $ \left(f_{Dd}(bb)\right)^{2}=0.5$, we get

\begin{eqnarray}
    \Gamma\left(T_{bb\bar{u}\bar{d}}\left(1^+\right)\to T_{bc\bar{u}\bar{d}}\left(0^+\right)\pi^-\right)&=&0.23\; \frac{\left(f_{Dd}(bb)\right)^{2}} {0.5}\times 10^{-12}\; \text{MeV}.
    \label{Ndrate-pi}
\end{eqnarray}
Similarly for the $1^+\to 1^+$ case, the result from Eq. (\ref{drate-11pi}) is:
\begin{equation}
  \Gamma\left(T_{bb\bar{u}\bar{d}}\left(1^+\right)\to T_{bc\bar{u}\bar{d}}\left(1^+\right)\pi^-\right)=1.16\; \frac{\left(f_{Dd}(bb)\right)^{2}} {0.5}\times 10^{-12}\;\text{MeV}.\label{Ndrate1p-pi}   
\end{equation}
For the decays $T_{bb\bar{u}\bar{d}}\left(1^+\right)\to T_{bc\bar{u}\bar{d}}\left(0^+ , 1^+ \right) D^{-}_s$ (see Fig.~\ref{FeynDiag2}(c)), 
 replacing $V_{ud}$ with $V_{cs}$, and using the values of the masses and the decay constant from Table \ref{tab:MesonPar},
 w e get the following partial widths
\begin{eqnarray}
 \Gamma\left(T_{bb\bar{u}\bar{d}}\left(1^+\right)\to T_{bc\bar{u}\bar{d}}\left(0^+\right)D^{-}_s\right)&=&0.53\; \frac{\left(f_{Dd}(bb)\right)^{2}} {0.5}\times 10^{-12}\; \text{MeV}\notag\\
 \Gamma\left(T_{bb\bar{u}\bar{d}}\left(1^+\right)\to T_{bc\bar{u}\bar{d}}\left(1^+\right)D^{-}_s\right)&=&3.45\; \frac{\left(f_{Dd}(bb)\right)^{2}} {0.5}\times 10^{-12}\;\text{MeV}.\label{Ndrate1p-Ds}
 \end{eqnarray}
In the case of a vector or an axial-vector meson in the final state, the expressions for the decay widths are:
\begin{eqnarray}
\Gamma^{1^+\to 0^+} &=&  \frac{G_F^2\left|V_{cb}V_{ud}^*\right|^2 a_1^2 (\mu) f_{V}^2}{192\pi m_{T}^{4}m_{T^\prime}}\left|\xi_{T\to T^\prime}\left(w\right)\right|^2 \lambda^{1/2} \left(m_{T}^2,m_{T^\prime}^2,m_{V}^2\right) \left(\left(m_{T}+m_{T^\prime}\right)^2-m_{V}^2\right)\notag\\
&& \times\left(\left(m_{T}^2+m_{T^\prime}^2\right)^2-4m_{V}^4+m_{V}^2\left(3m_{T}^2+2m_{T}m_{T^\prime}+3m_{T^\prime}^2\right)\right),\label{drate-1zrho} 
\end{eqnarray}
\begin{eqnarray}
 \Gamma^{1^+\to 1^+} &=&  \frac{G_F^2\left|V_{cb}V_{ud}^*\right|^2 a_1^2 (\mu) f_{V}^2}{192\pi m_{T}^{4}m_{T^\prime}}\left|\xi_{T\to T^\prime}\left(w\right)\right|^2 \lambda^{1/2}\left(m_{T}^2,m_{T^\prime}^2,m_{V}^2\right) \left(\left(m_{T}+m_{T^\prime}\right)^2-m_{V}^2\right)\notag\\
&& \times \left(5\left(m_{T}^2+m_{T^\prime}^2\right)^2-8m_{V}^4+m_{V}^2\left(3m_{T}^2-2m_{T}m_{T^\prime}+3m_{T^\prime}^2\right)\right),
\label{drate-1p1rho} 
\end{eqnarray}
where the subscript $V$ defines $\rho^-$ or $a^-_1$. Using the numerical values from Table \ref{tab:MesonPar}, we get the
partial decay rates for the $1^+ \to 0^+$ transitions
\begin{eqnarray}
    \Gamma\left(T_{bb\bar{u}\bar{d}}\left(1^+\right)\to T_{bc\bar{u}\bar{d}}\left(0^+\right)\rho^-\right)&=&0.66\;\frac{\left(f_{Dd}(bb)\right)^{2}} {0.5}\times 10^{-12}\; \text{MeV}, \notag\\
    \Gamma\left(T_{bb\bar{u}\bar{d}}\left(1^+\right)\to T_{bc\bar{u}\bar{d}}\left(0^+\right)a_{1}^-\right)&=&0.91\;\frac{\left(f_{Dd}(bb)\right)^{2}} {0.5}\times 10^{-12}\; \text{MeV}. 
     \label{Ndrate1pz-rhoa}
\end{eqnarray}

The numerical values of the decay rates for the $1^+ \to 1^+$
transitions are:
\begin{eqnarray}
     \Gamma\left(T_{bb\bar{u}\bar{d}}\left(1^+\right)\to T_{bc\bar{u}\bar{d}}\left(1^+\right)\rho^-\right)&=&3.0\;\frac{\left(f_{Dd}(bb)\right)^{2}} {0.5}\times 10^{-12}\; \text{MeV},\notag\\
     \Gamma\left(T_{bb\bar{u}\bar{d}}\left(1^+\right)\to T_{bc\bar{u}\bar{d}}\left(1^+\right)a_1^-\right)&=&3.7\;\frac{\left(f_{Dd}(bb)\right)^{2}} {0.5}\times 10^{-12}\; \text{MeV}.\label{Ndrate1p1-rho}
     \end{eqnarray}
     For the decays $T_{bb\bar{u}\bar{d}}\left(1^+\right)\to T_{bc\bar{u}\bar{d}}\left(0^+ , 1^+ \right) D^{*-}_s$, the partial decay widths are 
  calculated to have the numerical values
\begin{eqnarray}
     \Gamma\left(T_{bb\bar{u}\bar{d}}\left(1^+\right)\to T_{bc\bar{u}\bar{d}}\left(0^+\right)D^{*-}_{s}\right)&=&2.1\;\frac{\left(f_{Dd}(bb)\right)^{2}} {0.5}\times 10^{-12}\; \text{MeV},\notag\\
     \Gamma\left(T_{bb\bar{u}\bar{d}}\left(1^+\right)\to T_{bc\bar{u}\bar{d}}\left(1^+\right)D^{*-}_{s}\right)&=&6.3\;\frac{\left(f_{Dd}(bb)\right)^{2}} {0.5}\times 10^{-12}\; \text{MeV}.\label{Ndrate1p1-Dss}
     \end{eqnarray}
The resulting branching ratios for the decays
$T_{bb\bar{u}\bar{d}}\left(1^+\right)\to T_{bc\bar{u}\bar{d}}\left(0^+\right)(\pi^-, \rho^-,a_1^-,D^{-}_s,D^{*-}_{s}) $ and
$T_{bb\bar{u}\bar{d}}\left(1^+\right)\to T_{bc\bar{u}\bar{d}}\left(1^+\right)(\pi^-, \rho^-,a_1^-,D^{-}_s,D^{*-}_{s}) $
are collected in Table~\ref{tab:brachratio}. As for the semileptonic branching ratios, we
use the total width $\Gamma_{\rm total}(T_{bb\bar{u}\bar{d}})=8.2 \times 10^{-10}$ MeV,  derived from the  lifetime $\tau(T_{bb\bar{u}\bar{d}})=0.8$~ps~\cite{Ali:2018ifm},
Since the $J^P=1^+$ tetraquark
$T_{bc\bar{u}\bar{d}}\left(1^+\right)$ is expected
to decay radiatively to the $J^P=0^+$ state, the
branching ratios for $T_{bb\bar{u}\bar{d}}\left(1^+\right)\to T_{bc\bar{u}\bar{d}}\left(0^+\right)(\pi^-, \rho^-,a_1^-, D_s^-, D^{*-}_s) +(\gamma)$ may reach
about $3$\%.

\begin{table}[tbp]
\scalebox{1.01}{
\renewcommand{\arraystretch}{1.5}
\begin{tabular}{|c|c|c|c|c|c|c|}
\hline\hline
Decay Process & $T_{bb\bar{u}\bar{d}}\to T_{bc\bar{u}\bar{d}}\;\ell^-\nu_{\ell}$ & $T_{bb\bar{u}\bar{d}}\to T_{bc\bar{u}\bar{d}}\;\tau^-\nu_{\tau}$ & $T_{bb\bar{u}\bar{d}}\to T_{bc\bar{u}\bar{d}}\;\pi^-$ & $T_{bb\bar{u}\bar{d}}\to T_{bc\bar{u}\bar{d}}\;\rho^-\left(a_{1}^-\right)$ & $T_{bb\bar{u}\bar{d}}\to T_{bc\bar{u}\bar{d}}\;D^{-}_s\left(D^{*-}_s\right)$ \\ \hline
$\mathcal{B}\left(1^+\to 0^{+}\right)$ & $4.0$ & $1.4$ & $0.3$ & $0.8\left(1.1\right)$ & $0.64\left(2.5\right)$ \\
\hline
Decay Process & $T_{bb\bar{u}\bar{d}}\to T_{bc\bar{u}\bar{d}}\;\ell^{-}\nu_{\ell}$ & $T_{bb\bar{u}\bar{d}}\to T_{bc\bar{u}\bar{d}}\;\tau^{-}\nu_{\tau}$ & $T_{bb\bar{u}\bar{d}}\to T_{bc\bar{u}\bar{d}}\;\pi^-$ & $T_{bb\bar{u}\bar{d}}\to T_{bc\bar{u}\bar{d}}\;\rho^-\left(a_{1}^-\right)$ & $T_{bb\bar{u}\bar{d}}\to T_{bc\bar{u}\bar{d}}\;D^{-}_s\left(D^{*-}_s\right)$ \\ \hline
$\mathcal{B}\left(1^+\to 1^{+}\right)$  & $12.0$ & $5.3$ & $1.4$ & $3.7\left(4.5\right)$ & $4.2\left(7.7\right)$ \\
\hline
Decay Process & $T_{bc\bar{u}\bar{d}}\to T_{cc\bar{u}\bar{d}}\;\ell^{-}\nu_{\ell}$ & $T_{bc\bar{u}\bar{d}}\to T_{cc\bar{u}\bar{d}}\;\tau^-\nu_{\tau}$ & $T_{bc\bar{u}\bar{d}}\to T_{cc\bar{u}\bar{d}}\;\pi^-$ & $T_{bc\bar{u}\bar{d}}\to T_{cc\bar{u}\bar{d}}\;\rho^-\left(a_{1}^-\right)$ & $T_{bc\bar{u}\bar{d}}\to T_{cc\bar{u}\bar{d}}\;D^{-}_s\left(D^{*-}_s\right)$ \\ \hline
$\mathcal{B}\left(0^+\to 1^{+}\right)$  & $3.5$ & $1.1$ & $0.25$ & $0.7\left(1.0\right)$ & $5.8\left(2.4\right)$\\
\hline\hline
\end{tabular}%
}
\caption{Semileptonic and non-leptonic branching ratios (in units of $10^{-3}$) of the doubly-bottom
$T_{bb \bar u \bar d} $ and bottom-charm 
$T_{bc \bar u \bar d}$ tetraquarks with 
the indicated spin-parity. Note that we have set $\left(f_{Dd}(bb)\right)^{2} =0.5$ for the   $T_{bb\bar{u}\bar{d}}$ decays and $\left(f_{Dd}(bc)\right)^{2} =0.5$ for the $T_{bc\bar{u}\bar{d}}$ decays.}
\label{tab:brachratio}
\end{table}

\section{Semileptonic and non-eptonic decays $T_{bc\bar{u}\bar{d}}\left(J^P=0^{+}\right)\to T_{cc\bar{u}\bar{d}} \left(J^P=1^+\right) (\ell^-\nu_\ell, h^-)$}\label{sec:5}

The decay $T_{bc\bar{u}\bar{d}}\to T_{cc\bar{u}\bar{d}}\ell^{-}\nu_{\ell}$ is governed by the
effective Hamiltonian given in Eq.~(\ref{eq:06}), and the corresponding Feynman diagram is drawn in Fig.~\ref{FeynDiag1}(b).
We identify the tetraquark $T_{cc\bar{u}\bar{d}}$ with the
doubly-charm hadron $T_{cc} (3875)^+$, discovered by the LHCb,
having $J^P=1^+$~\cite{LHCb:2021vvq}, which implies that 
we have a scalar to axial-vector transition. The HQET matrix element in this case is parameterized as
\begin{align}
\frac{\left\langle T_{cc\bar{u}\bar{d}}\left(v^{\prime},\varepsilon^{\prime}\right)\left|\bar{c}\gamma^{\mu}b\right|T_{bc\bar{u}\bar{d}}\left(v\right)\right\rangle }{\sqrt{m_{T}m_{T^{\prime}}}} & = h_1 \left(w\right)i\varepsilon^{\mu\nu\alpha\beta}\varepsilon_{\nu}^\prime v_{\alpha}v_{\beta}^{\prime}, \nonumber \\
\frac{\left\langle T_{cc\bar{u}\bar{d}}\left(v^{\prime},\varepsilon^\prime\right)\left|\bar{c}\gamma^{\mu}\gamma_{5}b\right|T_{bb\bar{u}\bar{d}}\left(v\right)\right\rangle }{\sqrt{m_{T}m_{T^{\prime}}}} & = h_2 \left(w\right)\left(w+1\right)\varepsilon^{\prime\mu}-h_{3}\left(w\right)\left(\varepsilon^\prime\cdot v \right)v^{\prime\mu}-h_{4}\left(w\right)\left(\varepsilon^{\prime} \cdot v \right)v^{\mu}. 
\label{eq:bc1}
\end{align}
The spin symmetry relating the form factors in this case becomes
\begin{equation}
\left\langle \mathcal{H}_{A, v^\prime}^{\left(Q^{\prime}\right)}\left|\bar{Q}^{\prime}\Gamma Q\right|\mathcal{H}_{S, v}^{\left(Q\right)}\right\rangle =-\xi\left(w\right)\text{Tr}\left\{ \bar{\mathcal{H}}_{A, v^\prime}^{\left(Q^{\prime}\right)}\Gamma\mathcal{H}_{S, v}^{\left(Q\right)}\right\} ,\label{eq:bc2}
\end{equation}
and by using Eqs.~(\ref{eq:04}) and~(\ref{eq:07}) in Eq.~(\ref{eq:bc2}),
the matrix elements for vector and axial-vector currents become
\begin{align}
\left\langle \mathcal{H}_{A, v^\prime}^{\left(Q^{\prime}\right)}\left|\bar{c} \gamma^{\mu}b \right|\mathcal{H}_{S, v}^{\left(Q\right)}\right\rangle  & =-\xi\left(w\right)\text{Tr}\left\{ P_{v^\prime}^{* \left(Q^\prime\right) \dagger \alpha}\gamma_{\alpha}\gamma_{5}\frac{1+\slashed{v}^\prime}{2}\gamma^{\mu}\frac{1+\slashed{v}}{2}P_{v}^{*\left(Q\right)}\right\} \nonumber \\
 & =i\xi\left(w\right)\varepsilon^{\mu\nu\alpha\beta}\varepsilon^{\prime}_{\nu}v_{\alpha}v_{\beta}^{\prime}, \nonumber \\
\left\langle \mathcal{H}_{A, v^\prime}^{\left(Q^{\prime}\right)}\left|\bar{c} \gamma^{\mu}b \right|\mathcal{H}_{S, v}^{\left(Q\right)}\right\rangle   & =-\xi\left(w\right)\text{Tr}\left\{ P_{v^\prime}^{*\alpha\dagger\left(Q^\prime\right)}\gamma_{\alpha}\gamma_{5}\frac{1+\slashed{v}^\prime}{2}\gamma^{\mu}\gamma_{5}\frac{1+\slashed{v}}{2}P_{v}^{*\left(Q\right)}\right\}\nonumber \\
 & =\xi\left(w\right)\left[\left(w+1\right)\varepsilon^{\prime\mu}-\left(\varepsilon^\prime \cdot v\right)v^{\prime \mu}\right].
\label{eq:bc3}
\end{align}

Comparing Eq. (\ref{eq:bc2}) and Eq. (\ref{eq:bc3}),  one gets
\begin{equation}
h_1 \left ( w \right ) = h_2 \left ( w \right ) = h_3 \left ( w \right ) = \xi \left ( w \right ) , 
\quad 
h_4 \left ( w \right ) = 0.
\label{eq:10-2}
\end{equation}


Therefore, the hadronic transition amplitude in this case becomes
%
\begin{align}
\mathcal{M}_{T_{bc}\to T_{cc}}^{\mu} & =-\frac{G_{F}}{\sqrt{2}}V_{cb}\xi_{T\to T^{\prime}}\left(w\right)\sqrt{m_{T}m_{T^{\prime}}}\left[\left(w+1\right)\varepsilon^{\prime\mu}-\left(\varepsilon^\prime\cdot v\right)v^{\prime \mu}-i\varepsilon^{\mu\nu\alpha\beta}\varepsilon^\prime_{\nu}v_{\alpha}v_{\beta}^{\prime}\right].\label{ampbc2}
\end{align}
The relation between the form factors given in Eq.~(\ref{eq:12}) holds in this case as well.  
Contracting the hadronic part with the leptonic part given in Eq.~(\ref{eq:lep}), and summing over the polarization, i.e., $\sum_{\lambda} \varepsilon^{\prime}_{\mu}\left(v^\prime,\lambda\right)\varepsilon_{\nu}^{\prime*}\left(v^\prime,\lambda\right) = -g_{\mu\nu}+v^{\prime}_\mu v^{\prime}_\nu$, it is easy to see that the expression for the decay rate is again given by Eq.~(\ref{drate-mell}). However, in this case the initial state tetraquark has $J^{P} = 0^{+}$, therefore, $g = 1$. 

\vspace*{3mm}
Using the numerical values of the various input parameters from Table~\ref{tab:MesonPar}, the total decay width $\Gamma_{\rm total}(T_{bc\bar{u}\bar{d}})=2.2 \times 10^{-9}$~MeV, which is derived from the  lifetime $\tau(T_{bc\bar{u}\bar{d}})=0.3$~ps, based on the assumption that it is expected to be similar to the one calculated for the bottom-charm baryon $\tau(\Xi_{bc}^+ (bcu)) =(0.28 \pm 0.33)$~ps~\cite{Kiselev:2001fw},
together with $N_{b}=1$, and setting $\left(f_{Dd}(bc)\right)^2=0.5$,  we get the following branching ratios:
\begin{align}
 {\cal B} \left(T_{bc\bar{u}\bar{d}} (J^P=0^+) \to T_{cc\bar{u}\bar{d}} (J^P=1^+)\ell^{-}\nu_{\ell}\right) & =3.5\times\;\frac{\left(f_{Dd}(bc)\right)^{2}} {0.5} 10^{-3}~({\text{for}}~ \ell= e,\; \mu), \nonumber \\ 
 {\cal B} \left(T_{bb\bar{u}\bar{d}} (J^P=0^+) \to T_{bc\bar{u}\bar{d}} (J^P=1^+)\tau^{-}\nu_{\tau}\right) & =1.1\;\frac{\left(f_{Dd}(bc)\right)^{2}} {0.5} \times 10^{-3}. 
\label{bratios0to1}
\end{align}

In the case of the non-leptonic  decays, following the same procedure as adopted for the $T_{bb\bar{u}\bar{d}}\to T_{bc\bar{u}\bar{d}}$ case, the amplitude for  $h^- = \pi^{-}$ becomes
\begin{align}
\mathcal{M}^{0^{+}\to 1^{+}} & =-i\frac{G_{F}}{\sqrt{2}}V_{cb}V_{ud}^{*}a_{1}\left(\mu\right)f_{\pi}\xi_{T\to T^{\prime}}\left(w\right)q_{\mu}\left[\left(w+1\right)\varepsilon^{\prime\mu}-\left(\varepsilon^\prime \cdot v\right)v^{\prime\mu}-i\varepsilon^{\mu\nu\alpha\beta}\varepsilon^{\prime}_{\nu}v_{\alpha}v_{\beta}^{\prime}\right].\label{eq:bcpi}
\end{align}
From Eq. (\ref{2body-rate}), after performing the polarization sum, one gets the expression for the decay rate given in Eq. (\ref{drate-1zpi}) by setting
  $N_{b} = 1$, and $g=1$. Using the values of the input parameters from Table \ref{tab:MesonPar}, we get
\begin{equation}
    \Gamma\left(T_{bc\bar{u}\bar{d}}\left(0^+\right)\to T_{cc\bar{u}\bar{d}}\left(1^+\right)\pi^-\right)=0.55 \;\frac{\left(f_{Dd}(bc)\right)^{2}} {0.5} \times 10^{-12}\; \text{MeV}.
\end{equation}
Similarly, for the cases  $h^-=\rho^{-},\; a_{1}^{-}$, from Eq. (\ref{drate-1zrho}), the numerical values of the partial decay widths are:
\begin{eqnarray}
\Gamma\left(T_{bc\bar{u}\bar{d}}\left(0^+\right)\to T_{cc\bar{u}\bar{d}}\left(1^+\right)\rho^-\right)&=&1.55 \;\frac{\left(f_{Dd}(bc)\right)^{2}} {0.5} \times 10^{-12}\; \text{MeV}, 
\notag\\
\Gamma\left(T_{bc\bar{u}\bar{d}}\left(0^+\right)\to T_{cc\bar{u}\bar{d}}\left(1^+\right)a_1^-\right)&=&2.18 \;\frac{\left(f_{Dd}(bc)\right)^{2}} {0.5} \times 10^{-12}\; \text{MeV}.
\label{Ndrate2pz-rhoa}
\end{eqnarray}

For the decays $T_{bc\bar{u}\bar{d}}\left(0^+\right)\to T_{cc\bar{u}\bar{d}}\left(1^+ \right) (D_s^-, D^{*-}_s)$,
 the Feynman diagram is drawn in Fig.~\ref{FeynDiag2}(d). Using the numerical values of the input parameters from Table \ref{tab:MesonPar}, 
  the partial decay widths are:
\begin{eqnarray}
     \Gamma\left(T_{bc\bar{u}\bar{d}}\left(0^+\right)\to T_{cc\bar{u}\bar{d}}\left(1^+\right)D^{-}_s\right)&=&1.28 \;\frac{\left(f_{Dd}(bc)\right)^{2}} {0.5} \times 10^{-12}\; \text{MeV}\notag\\
    \Gamma\left(T_{bc\bar{u}\bar{d}}\left(0^+\right)\to T_{cc\bar{u}\bar{d}}\left(1^+\right)D^{*-}_s\right)&=&5.25 \;\frac{\left(f_{Dd}(bc)\right)^{2}} {0.5} \times 10^{-12}\; \text{MeV}.
     \label{Ndrate1pz-DsDss}
\end{eqnarray}
The branching ratios for $T_{bc\bar{u}\bar{d}}\left(0^+\right)\to T_{cc\bar{u}\bar{d}}\left(1^+\right)(\pi^-, \rho^-,a_1^-,D^{-}_s,D^{*-}_{s})$ are also collected
in Table \ref{tab:brachratio}. Together, they may reach about 1\%.
This completes the numerical part of this Letter.

\section{Summary and prospects of experimental measurements}\label{sec:6}
 Spectroscopic calculations  based
on Lattice QCD  support a substantial diquark-antidiquark component in the Fock space of the
tetraquarks $T_{bb\bar{u}\bar{d}}$ and
$T_{bb\bar{u}\bar{s}}$. Motivated by this input and the theoretical consensus that the
doubly-bottom tetraquarks $T_{bb\bar{u}\bar{d}}$ and
$T_{bb\bar{u}\bar{s}}$ are stable against strong and
electromagnetic interactions, we have worked out some
signature decays, reflecting the diquark-antidiquark component of
the $T_{bb\bar{u}\bar{d}}$ tetraquark wave-function.
They are characterised by the transitions 
$T_{bb\bar{u}\bar{d}} \to T_{bc\bar{u}\bar{d}} +X$,
where $X= \ell^- \nu_\ell$ or  charged mesons. 
We work out the branching ratios for the semileptonic decays for the assumed
$J^P$ quantum numbers of the initial and final state tetraquarks.
This requires the knowledge of the weak current matrix elements
(form factors) for the tetraquarks, which are not at hand. We use heavy quark
symmetry to relate the corresponding form factors at the symmetry
point and have argued that this should be a good approximation,
as the underlying weak Hamiltonian for the $B \to (D,D^*) \ell \nu_\ell$ and $T_{bb\bar{u}\bar{d}} \to T_{bc\bar{u}\bar{d}} 
\ell \nu_\ell$ is the same, resulting from the $b \to c W^-$
transition, involving heavy quarks in the initial and final states, and 
the momentum-transfer to the spectators is
 small in both cases.
A relation between the Isgur-Wise form factors of the doubly-bottom tetraquarks and the $B$-mesons follows in the symmetry limit. It is important to
calculate the form factors of the charged weak currents for the tetraquark to tetraquark transitions using
other theoretical techniques, in particular,  Lattice QCD:

\vspace*{3mm}
For the non-leptonic decays, we have concentrated for the cases, 
where $X$ is a single light meson $X=\pi^-, \rho^-, a_1^-$, resulting from $W^- \to \bar{u} d$, 
and $X=D_s^-, D^{*-}_s$, resulting from $W^- \to \bar{c} s$,
which represent the so-called  color-allowed  tree diagrams, and
have used factorization to write the hadronic matrix elements - a
method well-known from the decays $B \to (D,D^*) \pi$.
Following the Lattice-QCD study~\cite{Bicudo:2021qxj}, the tetraquark wave-function is composed of diquarkonic and mesonic components. In the present context, we
parametrize it by the diquarkonic fraction  
$\left(f_{Dd}(bb)\right)^2$, with the mesonic fraction given by $ 1- \left(f_{Dd}(bb)\right)^2$. 
   We use the Lattice-QCD suggested value $\left(f_{Dd}(bb)\right)^2=0.5$~\cite{Bicudo:2021qxj}, for the probability of finding the diquarkonic component in the $T_{bb\bar{u}\bar{d}}$ tetraquark, and the resulting
ratios are shown in Table \ref{tab:brachratio}. Some of them  are large enough to be measured at the HL-LHC, and all of them in the long run 
at the proposed Tera-$Z$ factories. We argue that experimental
measurements of some of the branching ratios presented here may also determine this fraction.
Since these decays involve the
bottom-charm tetraquarks in the final state, they have to be measured first in sufficient numbers. Detailed studies at the LHCb
are encouraging~\cite{Blusk:21,Polyakov:23}. There are good prospects at the two large experiments (ATLAS and CMS) at the LHC
due to their larger acceptance and much higher integrated luminosity. 
 Of course, many more decay modes of the tetraquarks following from the
$b$-quark decays $b \to c (\bar{u} d, \bar{c}s)$ can be calculated. 
In particular, they involve decays to doubly-heavy baryons, such as
$T_{bb\bar u \bar d} \to \Xi_{bc}^+ \;\bar{p} \;(\ell^- \nu_\ell)$ and $T_{bb\bar u \bar d} \to \Xi_{bc}^0 \;\bar{n} \;(\ell^- \nu_\ell)$.

\vspace*{3mm}
 We have also worked out the weak decays of the $J^P=0^+$ bottom-charm tetraquark 
 $T_{bc\bar{u}\bar{d}} \to T_{cc\bar{u}\bar{d}} +X$,
 where $X=\ell^- \nu_\ell$ or hadrons,
 with the final-state tetraquark identified with the observed doubly-charm narrow state 
 $T_{cc} (3875)^+$, having $J^P=1^+$, decaying to $DD^*$~\cite{LHCb:2021vvq}.
 Their branching ratios are also shown in Table \ref{tab:brachratio}, using $\left(f_{Dd}(bc)\right)^2=0.5$. Compared to the
corresponding decays of the doubly-bottom tetraquarks,
they are smaller due to the anticipated larger decay width of
the $T_{bc\bar{u}\bar{d}}$ tetraquark. We recall that
there are three branches of the weak decays of 
$T_{bc\bar{u}\bar{d}}$. Their relative fractions,
based on the anticipated similarities with the bottom-charm $\Xi_{bc}^+$-baryon decays~\cite{Kiselev:2001fw} are estimated as $b \to c$; $f_b=0.22 \pm 0.04$, $c \to s$; $f_c=0.72\pm 0.04$,
and $W^\pm$-exchange, $f_W=0.06\pm 0.04$~\cite{Polyakov:23}.
Of these, only the $b\to c$ transition will yield the
observed tetraquark $T_{cc} (3875)^+$. Other branches will lead to different final states, of which we expect
that the $c \to s$ transition may also lead to a tetraquark with the quark content $T_{bs \bar{u} \bar {d}}$ from the decays $T_{bc\bar{u}\bar{d}} \to T_{bs\bar{u}\bar{d}} +(\ell^+ \nu_\ell, X)$.
Due to the larger fraction of the $c \to s$ decay, $f_c$, this might be a promising place to discover a
tetraquark with four different quark flavors.

\vspace*{3mm}
In summary, weak decays of the doubly-heavy tetraquarks, some of which are discussed here, 
are anticipated to induce tetraquark $\to $ tetraquark transitions. 
This is due to the presence of a doubly-heavy diquark component in the Fock space of these hadrons. These signature decays of the compact tetraquarks  may be observed at the high-luminosity LHC experiments 
(LHCb, ATLAS and CMS)~\cite{Azzi:2019yne,Cerri:2018ypt,LHCb:2018roe}, and eventually at
the Tera-$Z$ factories, being considered at the future circular $e^+e^-$ machines at CERN~\cite{FCC:2018evy}
 and China~\cite{CEPCStudyGroup:2018ghi}.
Establishing these decays will provide a proof of the
existence of doubly-heavy diquarks, long anticipated
in the context of the doubly-heavy tetraquarks~\cite{Manohar:1992nd}, which have now received additional support from the Lattice-QCD studies.

\vspace*{-3mm}
\section*{Acknowledgements}
\vspace*{-3mm}
This project was initiated during the 5th International Workshop on Heavy Quark Physics at NCP,  
Islamabad. One of us (AA) would like to thank NCP for
the kind hospitality. We acknowledge helpful discussions with Timothy Gershon,
Luciano Maiani and Ivan Polyakov. We thank Alexander Parkhomenko for carefully reading the manuscript.

\end{document}